\documentclass{aa} 

\usepackage{psfig}

\begin{document}
\thesaurus{09.08.1, 10.19.2, 08.05.1, 08.06.2, 13.18.3}
\title{Young massive stars in the ISOGAL survey}
\subtitle{I. VLA observations of the ISOGAL $l$$=$$+$45 field}
\author{Leonardo Testi\inst{1,2}, Marcello Felli\inst{1}
and Gregory B. Taylor\inst{3}}
\offprints{Testi: Arcetri, lt@arcetri.astro.it}
\institute{
Osservatorio Astrofisico di Arcetri,
Largo E. Fermi 5, I-50125 Firenze, Italy
\and
Division of Physics, Mathematics and Astronomy,
California Institute of Technology, MS 105-24, Pasadena, CA 91125, USA
\and
National Radio Astronomy Observatory,
P.O. Box O, Socorro, NM 87801, USA}
\date{Received xxxx; accepted xxxx}
\maketitle
\markboth{L. Testi et al.: VLA observations of the ISOGAL $l$$=$$+$45 field}{L. Testi et al.: VLA observations of the ISOGAL $l$$=$$+$45 field}
\begin{abstract}

We present VLA radio continuum observations at 3.6 and 6~cm of a $\sim$0.65 sq.
deg. field in the galactic plane at $l=+45^{\circ}$. These observations
are meant to be used in a comparison with ISO observations at 7 and 15
$\mu$m of the same region. In this paper we compare the radio results with
other radio surveys and with the IRAS-PSC.

At 3.6 and/or 6~cm we detect a total of 34 discrete sources, 13 of
which are found in five separate extended complexes. These are all
multiple or single extended thermal ultra-compact HII (UCHII) regions.
While for each of these complexes an IRAS counterpart could be
reliably found, no IRAS counterpart could be reliably identified for
any of the remaining 21 sources.  Of these 21 compact sources, six are
candidate UCHII regions, and the other 15 are most probably
background extragalactic non-thermal sources.

The five IRAS sources associated with the radio continuum complexes all
satisfy the Wood \& Churchwell~(\cite{WC89}; WC89) color criteria for UCHII.
None of the other 38 IRAS point sources present in our surveyed field 
show the same colors. This fraction of WC89 type to total IRAS
sources is consistent with what is found
over the entire galactic plane. The fact that, when observed with a compact
VLA configuration, the IRAS sources with ``UCHII colors'' are found to 
be associated with arcminute-scale extended sources, rather than with 
compact or unresolved radio sources, may have important implications
on the estimated lifetime of UCHII regions.

\end{abstract}

\section{Introduction}
\label{sintro}

The present work is part of a larger project to study the galactic plane 
(ISOGAL, Omont \& Blommaert~\cite{O97}; 
P\'erault et al.~\cite{Pea96}). During the ISO mission, the 
ISOGAL consortium observed with ISOCAM
at 15 and 7 $\mu$m selected parts of the galactic plane (about 18 sq.deg.
distributed along the inner galactic disk) in order to
study the stellar populations in the inner galaxy,
with a sensitivity and resolution
two orders of magnitude better than IRAS.
The main scientific goal of the ISOGAL project was the study of
the distribution and properties of the AGB stars.
However, the survey is unbiased, with the only exception of excluding from the
surveyed area
strong IRAS sources (with 12 $\mu$m flux densities greater than 6-10
Jy) in order to avoid saturation effects. Thus the survey data
can be used to study any other type
of mid-IR source present in the galactic plane, as for instance the less
numerous HII regions associated to young massive stars.

For a proper identification of source types, the ISOGAL results 
need to be
compared with observations at other wavelengths.  In particular, for 
the study of AGB stars comparisons with near IR 
observations, taken primarily with DENIS (Epchtein~\cite{E98}), 
are useful.  For the  study of HII regions comparisons
with radio continuum surveys are more appropriate.

A large fraction of the northern sky galactic fields covered by ISOGAL
have already been observed at 6~cm (5~GHz)
with the VLA (see Becker et al 1994 and references 
therein), and a comparison of the two surveys is underway.  
However, these radio observations 
terminate at $l=+40^{\circ}$
and there were no high frequency (e.g $\ge$5~GHz) radio 
continuum observations for the ISOGAL field at $l=+45^{\circ}$. 
Observations at lower frequencies, such as the 1.4 GHz Northern
VLA Sky Survey (NVSS -- Condon et al.~\cite{Cea98}),  are inadequate to detect 
the younger and more dense compact HII regions, which
may be optically thick at 1.4 GHz.

Given our interest, within the ISOGAL team, to study the young massive
stars, we decided to observe the $l=+45^{\circ}$ field at high frequencies 
with the VLA, to provide a data base comparable to that of Becker et al.\ (1994).
In order to obtain radio spectral index information 
we covered at 6 and 3.6~cm an area slightly larger than 
the $l=+45^{\circ}$ ISOGAL field.

The selection of the ISOGAL galactic  plane
fields does not follow any {\it ad hoc} criterion,
but is based on symmetrically spaced samples on both sides of the Galactic Center,
with the spacing increasing with distance from the Galactic Center.
The $l=+45^{\circ}$ field happens to be located tangent to a spiral
arm of our Galaxy, the Scutum arm (see e.g. Kurtz et al. 1994). Inspection
of the 4.875 GHz galactic plane survey of Altenhoff et al.~(\cite{Aea78})
shows that there is very weak diffuse galactic background emission
in this direction. Only 7 sources of the Altenhoff et al. catalogue fall
in our surveyed area or at its borders (see Table~\ref{talt}).
One of these (44.786--0.490) is partly outside our surveyed area. 
Most of these sources are associated with bright IRAS point sources
and have not been covered by the ISOCAM observations except for 
45.202--0.441 and 45.341--0.370.

In this work we present the radio observations and discuss the
comparison with other radio surveys and with IRAS data. Comparison
with ISOGAL data, as well as with dedicated J, H, K observations of 
the same field taken with TIRGO 
will be the subject of following works.

\section{Observations and data reduction}
\label{sobs}

The ISOGAL field centered at $l=+45^\circ$, $b=0^\circ$ was observed 
at 6 (4.9~GHz) and 3.6~cm (8.5~GHz) 
using the NRAO\footnote{The National Radio Astronomy Observatory is a 
facility of the National Science Foundation operated under agreement by the
Associated Universities, Inc.} Very Large Array (VLA) in the  C
configuration on 1997 August 5 (8~hours).
At 6~cm the observational setup was similar to that used by
Becker et al.~(\cite{Bea94}), the only differences being that 
our pointing centers are more closely packed and, due to the
peculiar geometry of the sub-fields observed with ISO, we covered the 
field by scanning strips at constant galactic longitude, which required
a total of 31 pointings;
our integration time per position was 300~s. At 3.6~cm we used a similar
pointing scheme but scaled due to the smaller primary beam. The observing
time per position was reduced to 210~s, and the entire field was mapped
with 74 pointings. 8 pointings were observed at 3.6~cm during a 1~hour test
run on 1997 July~4, however, due to a bug in the schedule, only some of the
pointings fell in our survey region. For the sake of completeness we will
also report the results for the 3 pointings outside our formal survey region
that form a spur in position angle 30 degrees.

\begin{figure}
\centerline{\psfig{figure=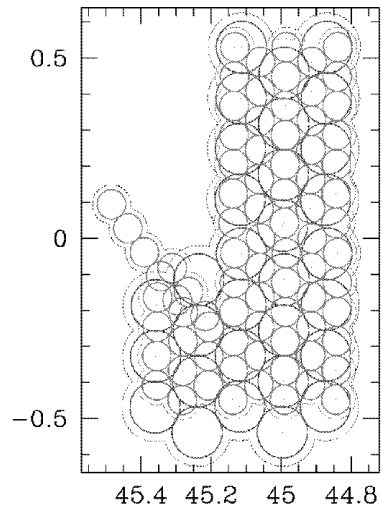,height=7cm}}
\caption[]{\label{fcover} At each pointing position a circle with diameter
equal to the VLA primary beam FWHM is shown. Grey circles represent 3.6~cm
pointings, black circles 6~cm pointings. The dotted line marks the 
boundaries of the area observed at both frequencies, the dashed line
encompasses the area observed at either of the two bands. In both cases
we considered only the area where the a primary beam attenuation is
less than a factor of 3. Axes are galactic longitude and latitude (degrees).}
\end{figure}

Due to the ill-determined primary beam correction and the rapid loss of
sensitivity far from the pointing center, we searched for sources only 
the area where the primary beam attenuation is less than a factor of 3.
With this constraint, we covered an area of $\sim$0.620~sq.~deg. at 6~cm,
and $\sim$0.525~sq.~deg. at 3.6~cm.
In Fig.~\ref{fcover} we show all the pointing positions:  the small grey circles
represent the VLA primary beam HPBW at 3.6~cm (4.9$^\prime$), while the
larger black circles those at 6~cm (8.6$^\prime$).
The dotted line show the boundaries of the area covered at both wavelengths
($\sim$0.493~sq.~deg.), the ISOGAL sub-fields are included in
this area, the dashed lines mark the boundary
of the field observed either at 6 and/or 3.6~cm ($\sim$0.652~sq.~deg.).

\begin{figure*}
\centerline{\psfig{figure=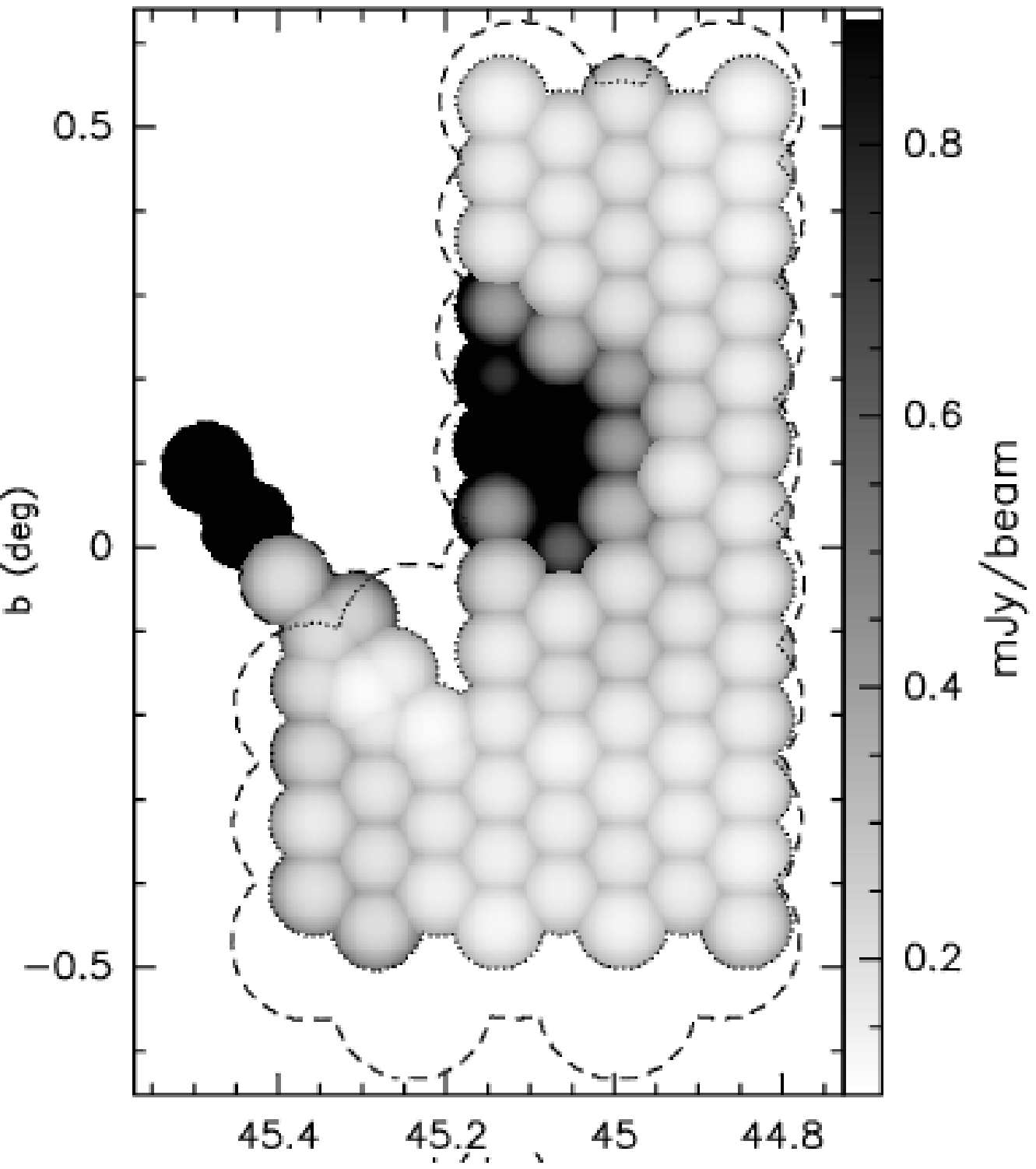,height=9cm,angle=-90}
            \hskip 0.4cm
            \hskip 0.4cm
            \psfig{figure=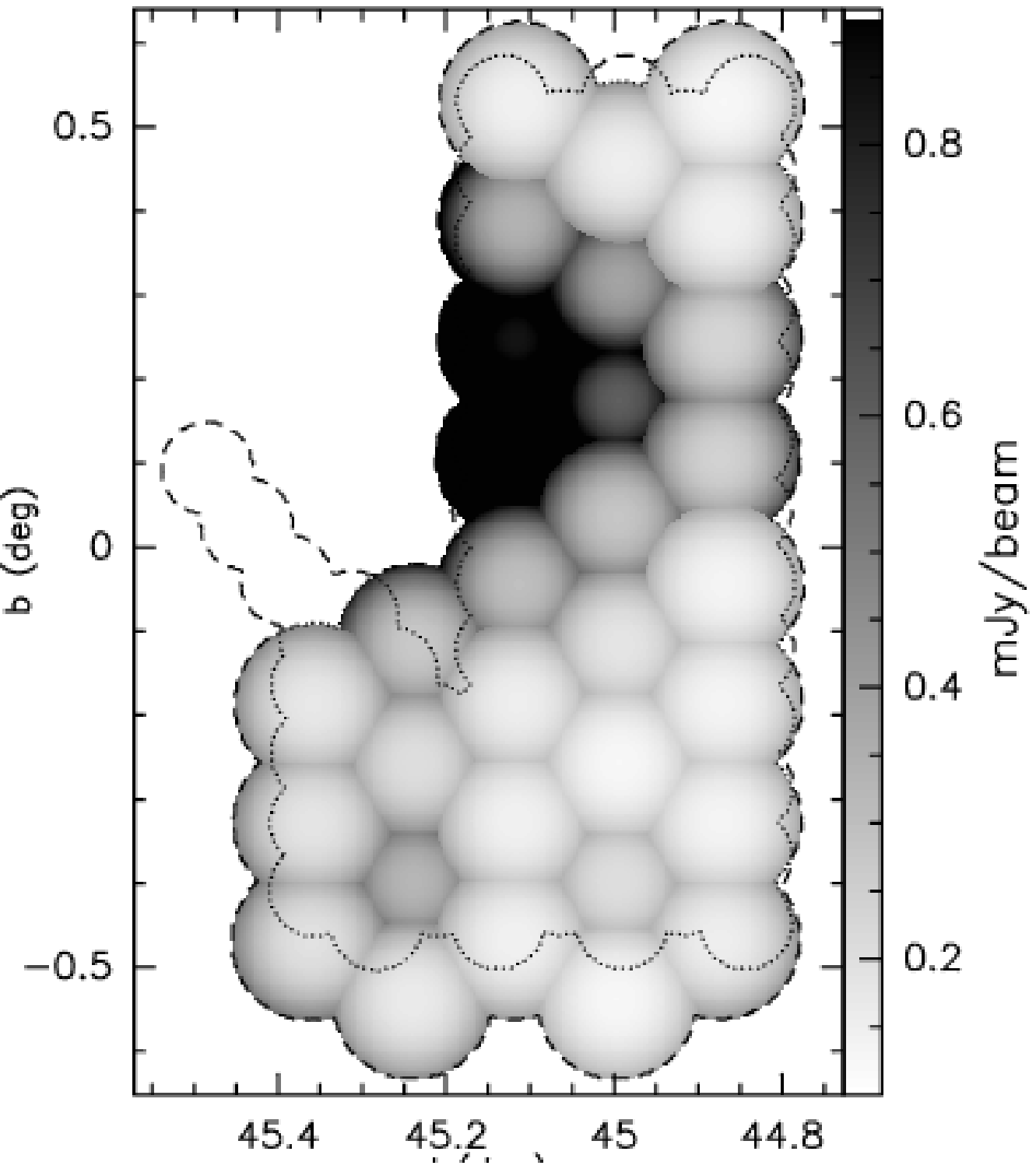,height=9cm,angle=-90}}
            \vskip 0.4cm
\caption[]{\label{frms}Computed noise maps for the 3.6 and 6~cm observations 
(left and right, respectively). Dotted and dashed lines as in Fig.~\ref{fcover}.
The on-axis noise level in the black areas can be as high as 8~mJy/beam.}
\end{figure*}

Frequent observations of the quasar 1922$+$155 were used for gain and phase
calibration, while the flux density scale was determined by observations
of 3C286. The calibration is expected to be accurate within 10\%.
We imaged all the fields using the AIPS IMAGR task with natural weighting,
the resulting synthesized beam varied somewhat from field to field
depending on the hour angle at which each field was observed,
typical FWHM values are $\sim 6^{\prime\prime}$
at 6~cm and $\sim 3^{\prime\prime}$ at 3.6~cm.


\subsection{Sensitivity}

Due to the VLA primary beam attenuation and the different noise values
in the various fields, the sensitivity of our observations is not 
expected to be uniform accross the observed region.
Using our knowledge of the VLA primary beam attenuation pattern and 
the measured on-axis rms level in each of the observed fields, we 
computed the sensitivity maps for our survey at 3.6 and 6~cm
(see also Zoonematkermani et al.~\cite{Zea90} and Becker et al.~\cite{Bea94}).
The measured on-axis noise level in the maps is generally
$\sim 0.12$--$0.15$~mJy/beam at both frequencies, with the exception
of some fields close to the bright complexes located at 
$l\sim 45^\circ\!.10$, $b\sim0^\circ\!.13$
($\alpha(2000)=19^h13^m27^s$
$\delta(2000)=10^\circ 53^\prime35^{\prime\prime}$)
and  $l\sim 45^\circ\!.45$, $b\sim0^\circ\!.06$
($\alpha(2000)=19^h14^m21^s$
$\delta(2000)=11^\circ 09^\prime13^{\prime\prime}$)
which have a higher noise level (in the range 1--8~mJy/beam)
due to residual phase and amplitude errors.

\begin{figure}
\centerline{\psfig{figure=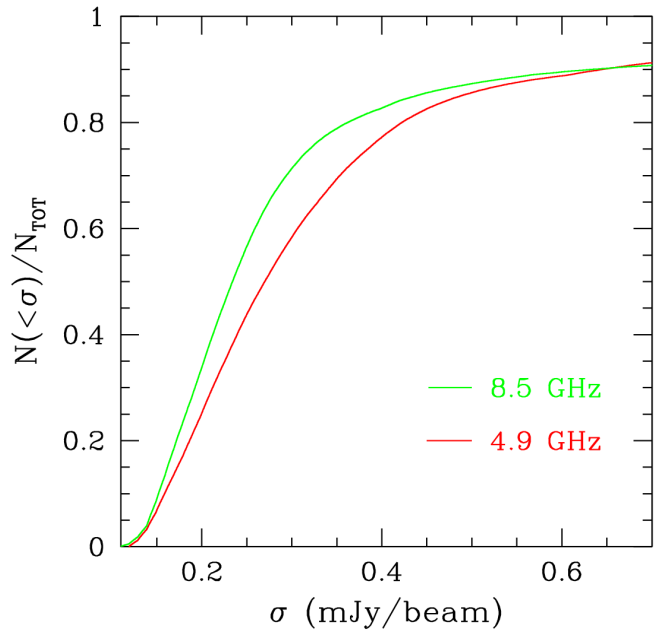,width=8.8cm}}
\caption[]{\label{frmscum}Cumulative distributions of the noise values in
the maps of Figure~\ref{frms}. The grey line is for 3.6~cm data,
the black line for 6~cm data.}
\end{figure}

The computed rms maps are shown in Fig.~\ref{frms}, the area of each pixel 
($10^{\prime\prime}\times 10^{\prime\prime}$) corresponds to $\sim$3.5 beams
at 6~cm and $\sim$14 beams at 3.6~cm. As seen from Fig.~\ref{frms}
most of the area covered by our survey has a rms sensitivity less 
than 0.3~mJy/beam at both frequencies. In Fig.~\ref{frmscum} we show the
cumulative distributions of the pixel values in the rms maps, 
more than 85\% of the surveyed area has an rms value less than 0.5~mJy/beam.

\subsection{Source extraction}

\begin{figure*}
\centerline{\psfig{figure=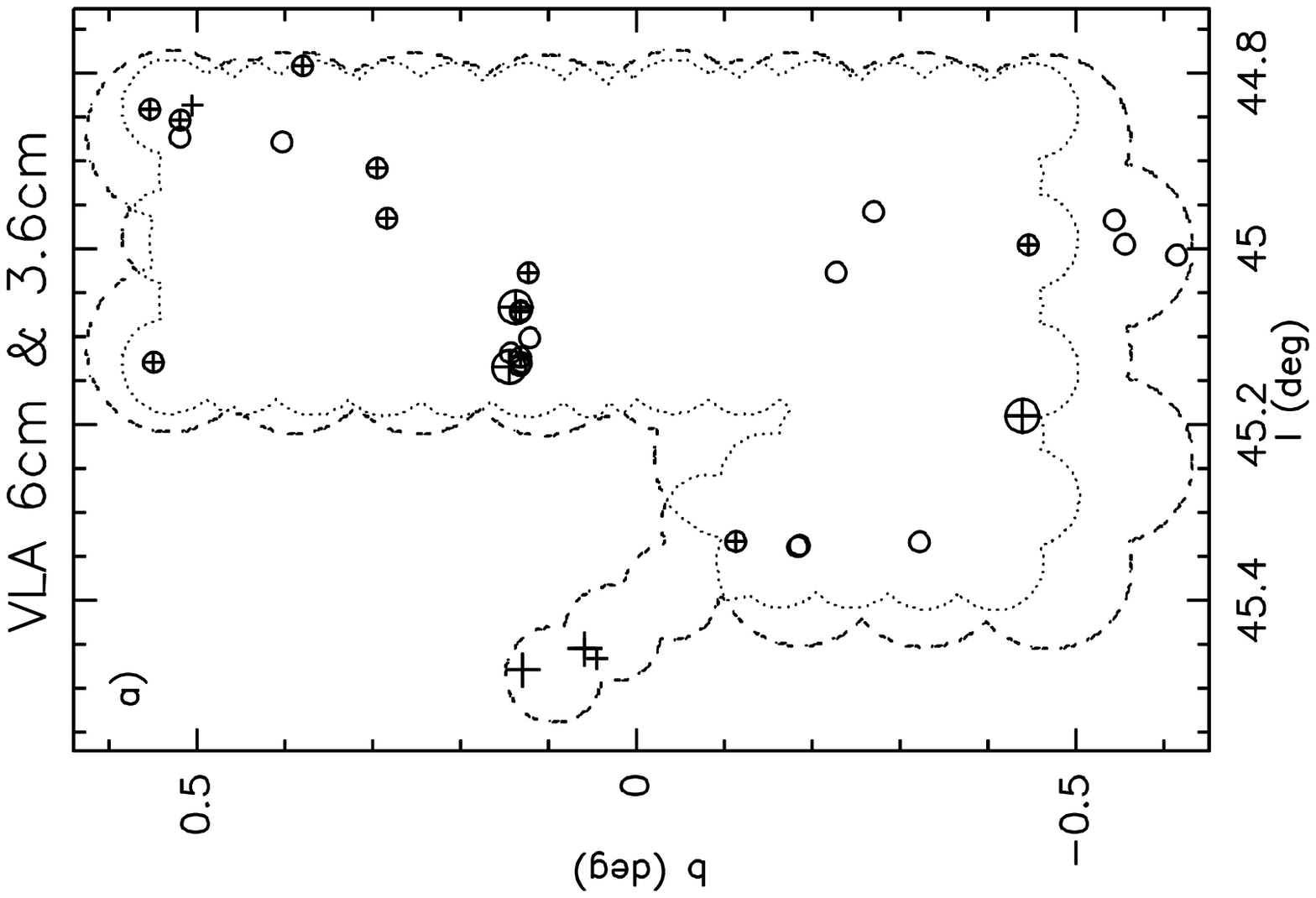,height=8.5cm,angle=-90}
            \psfig{figure=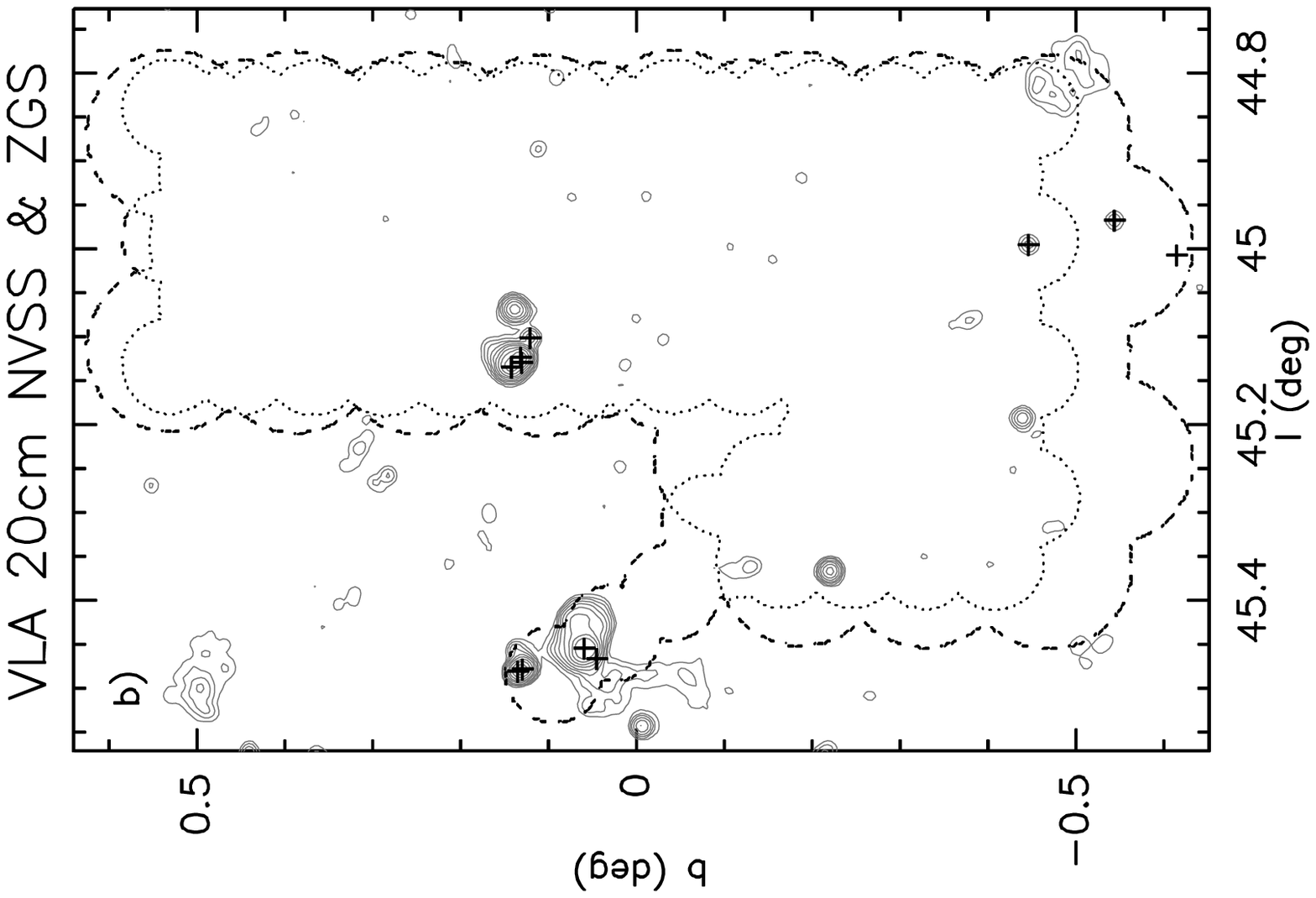,height=8.5cm,angle=-90}
            \psfig{figure=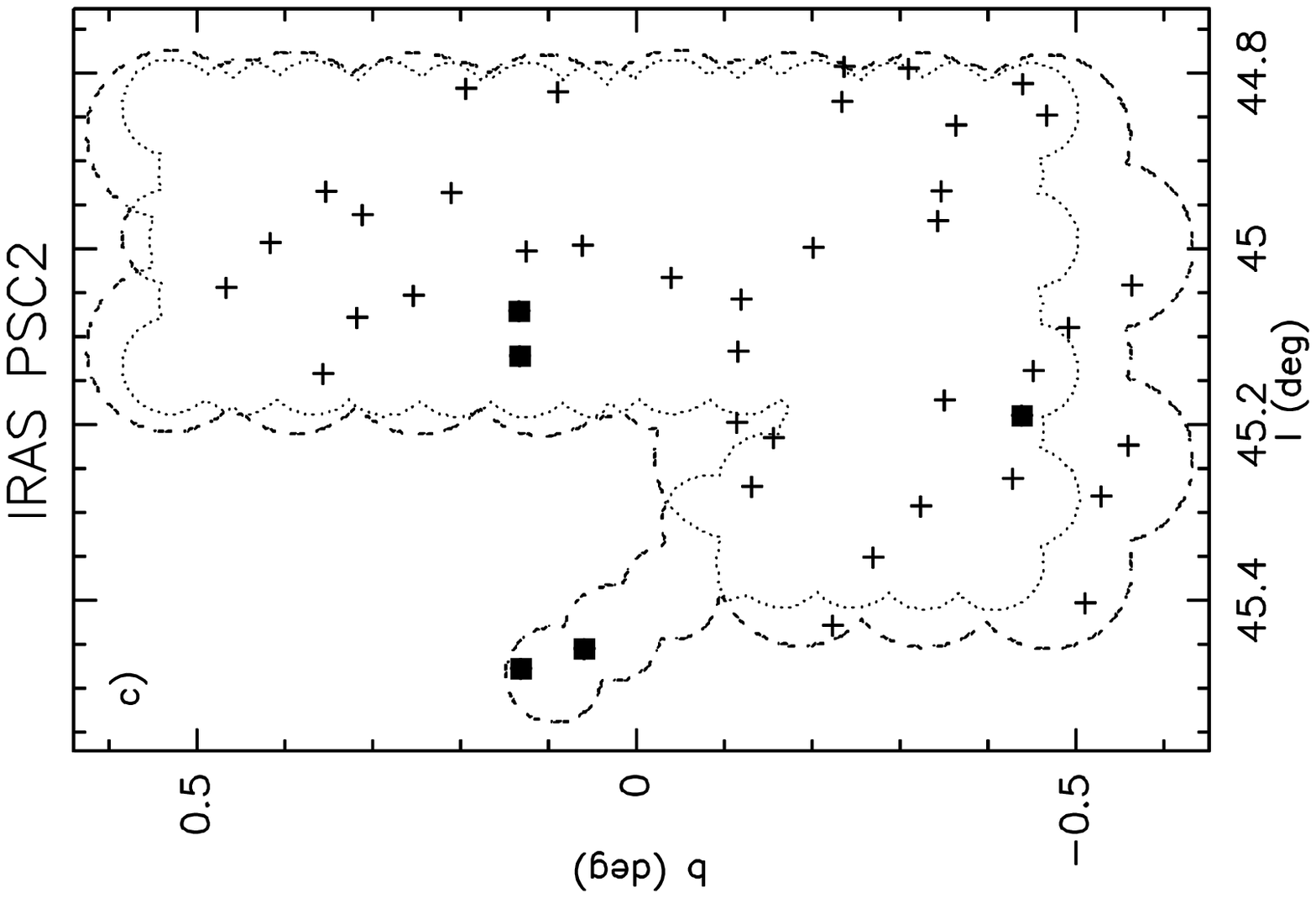,height=8.5cm,angle=-90}}
\caption[]{\label{fpos}
a) Positions of the detected sources at 3.6~cm (pluses) and 6~cm
(empty circles), larger symbols represent extended sources;
b) VLA 20~cm surveys:
sources from Zoonematkermani et al.~(\cite{Zea90}; ZGS) are shown as pluses, 
grey contours show the NVSS image of the region;
c) the position of the IRAS--PSC2 sources inside our extended survey
area are shown as plus symbols, filled squares show the five sources 
which satisfy the Wood \& Churchwell~(\cite{WC89}) color criteria.}
\end{figure*}

All images were inspected by means of contour plots and greyscale 
display to find sources. The images were then corrected for primary beam
attenuation using the AIPS task PBCOR before source extraction. 

The J2000.0 positions, peak and integrated flux densities and the sizes of the detected
sources at both frequencies are listed in Table~\ref{tsrc}.
In general, all the reported detections have a signal-to-noise 
ratio greater than five in at least one of the two bands.
The names assigned to the sources have been derived from their
galactic coordinates, as in Becker et al.~(\cite{Bea94}).

We arbitrarily divided the sources into two categories:
1) compact sources and 2) extended sources. In the first group
are all the unresolved sources or the sources with deconvolved sizes 
of the same order or smaller than the synthesized beam FWHM.
All the extended sources have sizes much greater than 
the synthesized FWHM, and thus they may be partially resolved
out by the interferometer. The flux densities (and sizes) reported in
Table~\ref{tsrc} for these sources should be considered as lower limits.
We shall see in the following that this arbitrary distinction 
based on observational considerations reflects also some 
intrinsic physical difference between the two groups.

At large distances from the pointing center, the correction factor due
to the VLA primary beam attenuation may be large, and hence the source
flux density could be ill-determined.
In Table~\ref{tsrc} the source that has the maximum correction factor applied
is source \#5, which is close to the edge of the 
surveyed area and has correction factors $\sim$2.1 at 6~cm and
$\sim$2.5 at 3.6~cm. All other sources, with the exception of \#22 and
\#29 at 6~cm, have correction factors lower than 2.0.

The positions of all the detected sources within our surveyed region
are shown in Fig.~\ref{fpos}~a), where pluses represent 3.6~cm sources
(21), and circles represent 6~cm sources (29), and larger symbols represent
extended sources.
Contour plots for all the detected sources are shown in the Appendix.

\begin{table*}
\caption[]{\label{tsrc}Detected radio sources}
\begin{tabular}{llccrrrrrrl}
\hline
  &   & & & \multicolumn{3}{c}{ -------------- 6~cm -------------- } & \multicolumn{3}{c}{ ------------ 3.6~cm ------------ } & \\
\# &Name$^a$& $\alpha$ & $\delta$ & \multicolumn{1}{c}{F$_p^b$} & \multicolumn{1}{c}{F$_i^b$} & \multicolumn{1}{c}{d$^b$} & \multicolumn{1}{c}{F$_p^b$} & \multicolumn{1}{c}{F$_i^b$} & \multicolumn{1}{c}{d$^b$} & Shown in\\
&&(2000)&(2000)&\multicolumn{1}{c}{(mJy/beam)}&\multicolumn{1}{c}{(mJy)}&\multicolumn{1}{c}{($^{\prime\prime}$)}&\multicolumn{1}{c}{(mJy/beam)}&\multicolumn{1}{c}{(mJy)}&\multicolumn{1}{c}{($^{\prime\prime}$)}&Figure$^c$\\
\hline
1&G044.841$+$0.554&19:11:24.65&$+$10:50:20.7&    1.4$\pm$   0.2&   1.6&0 &   0.8$\pm$   0.2&   1.1&1&\ref{pcfig1}\\
2&G044.854$+$0.519&19:11:33.51&$+$10:50:02.8&    0.7$\pm$   0.2&   0.7&1 &   0.6$\pm$   0.2&   1.6&3&\ref{pcfig1}\\
3&G044.837$+$0.506&19:11:34.52&$+$10:48:46.0&$<$   0.3& -- & -- &   0.7$\pm$   0.2&   0.7&0&\ref{pcfig2}\\
4&G044.873$+$0.520&19:11:35.64&$+$10:51:05.3&   0.9$\pm$   0.2&   1.4&5&$<$   0.6 & -- & -- &\ref{pcfig2}\\
5&G044.792$+$0.380&19:11:56.70&$+$10:42:53.5&    2.5$\pm$   0.2&   3.2&0 &   1.7$\pm$   0.2&   2.1&0&\ref{pcfig1}\\
6&G045.129$+$0.550&19:11:58.07&$+$11:05:32.9&    0.8$\pm$   0.2&   1.2&0 &   0.5$\pm$   0.2&   0.5&0&\ref{pcfig1}\\
7&G044.878$+$0.403&19:12:01.41&$+$10:48:09.0&   0.6$\pm$   0.2&   0.9&4&$<$   0.6 & -- & -- &\ref{pcfig2}\\
8&G044.908$+$0.295&19:12:28.20&$+$10:46:43.2&    1.5$\pm$   0.4&   1.6&1 &   2.0$\pm$   0.3&   2.5&1&\ref{pcfig1}\\
9&G044.965$+$0.284&19:12:37.07&$+$10:49:27.1&   18.4$\pm$   0.6&  20.1&1 &  18.5$\pm$   0.4&  18.1&0&\ref{pcfig1}\\
10&G045.027$+$0.123&19:13:18.94&$+$10:48:16.5&    9.0$\pm$   0.8&  10.1&0 &   7.2$\pm$   1.1&   6.6&0&\ref{pcfig1}\\
11&G045.070$+$0.132&19:13:21.87&$+$10:50:49.0&   63.2$\pm$   3.4& 270$^d$&6 &  57.6$\pm$   3.7& 106.6&2&\ref{exfig1}\\
12&G045.072$+$0.132&19:13:22.08&$+$10:50:53.2&  128.7$\pm$   3.6& 270$^d$&0 & 307.3$\pm$   3.9& 326.1&0&\ref{exfig1}\\
13&G045.118$+$0.143&19:13:24.90&$+$10:53:41.1&   24.0$\pm$   3.6& 172.0&6 &$<$24.0& -- & -- &\ref{exfig2}\\
14&G045.101$+$0.122&19:13:27.67&$+$10:52:09.6&  16.9$\pm$   3.4&  34.0&5&$<$   24.0 & -- & -- &\ref{pcfig2}\\
15&G045.123$+$0.132&19:13:27.91&$+$10:53:36.3& 1436.1$\pm$   3.4&2905.8&5 &1431.6$\pm$  17.2&3294.7&3&\ref{exfig2}\\
16&G045.133$+$0.133&19:13:28.81&$+$10:54:09.8&   24.0$\pm$   3.6& 88.0&4 &$<$24.0& -- & -- &\ref{exfig2}\\
17&G045.130$+$0.131&19:13:28.83&$+$10:53:56.1&   37.2$\pm$   3.6& 91.0&4 &38$\pm$8.0& 77 & 3 &\ref{exfig2}\\
18$^e$&G045.455$+$0.060&19:14:21.29&$+$11:09:12.3& --$^f$& --$^f$ & --$^f$ & 195.0$\pm$   4.2&1050.0$^e$&6&\ref{exfig3}\\
19&G045.466$+$0.045&19:14:25.66&$+$11:09:26.1& --$^f$& --$^f$ & --$^f$ &  87.2$\pm$   4.6& 105.4&1&\ref{pcfig2}\\
20&G045.026$-$0.227&19:14:34.64&$+$10:38:28.7&   0.9$\pm$   0.2&   0.8&0&$<$   0.6 & -- & -- &\ref{pcfig2}\\
21&G044.958$-$0.270&19:14:36.02&$+$10:33:37.7&   1.0$\pm$   0.2&   1.2&0&$<$   0.6 & -- & -- &\ref{pcfig2}\\
22&G045.333$-$0.113&19:14:44.71&$+$10:57:56.7&    4.1$\pm$   0.5&   4.4&0 &   3.6$\pm$   0.4&   4.3&1&\ref{pcfig1}\\
23&G045.339$-$0.183&19:15:00.62&$+$10:56:17.6&   2.1$\pm$   0.5&   3.3&3&$<$   0.9 & -- & -- &\ref{pcfig2}\\
24&G045.337$-$0.185&19:15:00.87&$+$10:56:09.7&   1.3$\pm$   0.5&   1.4&0&$<$   0.9 & -- & -- &\ref{pcfig2}\\
25&G044.996$-$0.446&19:15:18.35&$+$10:30:44.3&    7.5$\pm$   0.1&   7.8&0 &   4.2$\pm$   0.3&   4.6&0&\ref{pcfig1}\\
26&G045.333$-$0.322&19:15:29.98&$+$10:52:08.0&   1.3$\pm$   0.3&   2.3&3&$<$   0.6 & -- & -- &\ref{pcfig2}\\
27&G044.967$-$0.544&19:15:36.29&$+$10:26:30.6&   6.5$\pm$   0.2&   6.9&1& --$^f$ & --$^f$ & --$^f$ &\ref{pcfig2}\\
28&G044.995$-$0.555&19:15:41.95&$+$10:27:38.0&   1.1$\pm$   0.2&   1.0&0& --$^f$ & --$^f$ & --$^f$ &\ref{pcfig2}\\
29&G045.007$-$0.614&19:15:56.12&$+$10:26:38.1&   3.4$\pm$   0.2&   2.8&0& --$^f$ & --$^f$ & --$^f$ &\ref{pcfig2}\\
\hline
\multicolumn{9}{c}{Extended sources}\\
\hline
30&G045.066$+$0.138&19:13:20.5&$+$10:50:50&   39.5$\pm$   3.0& 348.6&26 &  14.9$\pm$   2.0& 433.0&26&\ref{exfig1}\\
31&G045.134$+$0.145&19:13:26.5&$+$10:54:20&   73.0$\pm$   3.0&1960.0&48 &  60.2$\pm$   8.0&1727.7&48&\ref{exfig2}\\
32&G045.479$+$0.130&19:14:08.8&$+$11:12:28& --$^f$& --$^f$ & --$^f$ & 37.2$\pm$  2.0&1500.0&30&\ref{exfig3}\\
33$^e$&G045.455$+$0.059&19:14:21.3&$+$11:09:10& --$^f$& --$^f$ & --$^f$ & 65.0$\pm$   2.0&3450.0$^e$&47&\ref{exfig3}\\
34&G045.190$-$0.439&19:15:39.0&$+$10:41:15&    7.6$\pm$   0.3&  95.6&36 &   3.5$\pm$   0.2&  69.7&36&\ref{exfig4}\\
\hline
\end{tabular}
\vskip 0.3cm
$^a$) Derived from galactic coordinates, as in Becker et al.~(\cite{Bea94})\\
$^b$) F$_p\equiv$ peak flux density; F$_i\equiv$ integrated flux density;
d$\equiv$ size (deconvolved).\\
$^c$) Contour plots for all the detected sources are reported in Appendix.\\
$^d$) Sources \#11 and \#12 are blended together at 6~cm, the separation 
of the two contribution to the integrated flux is very uncertain, thus we 
report the integrated flux of both components together..\\
$^e$) Source \#18 is inside the extended source \#33. The integrated
flux density of the compact component has been subtacted from the total 
integrated flux density, the resulting flux has been assigned to source \#33.\\
$^f$) Not observed.
\end{table*}

\section{Comparison with other surveys}
\label{sres}

\subsection{VLA 20~cm surveys}

The observed field is included in the VLA 20~cm galactic plane survey (ZGS;
Zoonematkermani~\cite{Zea90}) and in the NRAO-VLA Sky Survey
(NVSS; Condon et al.~\cite{Cea98}). Both these surveys used the 
VLA at 20~cm (1.4~GHz), however, the ZGS used the moderately extended B
array and has a typical peak flux density sensitivity of 25~mJy/beam
and a synthesized beam of $\sim 5^{\prime\prime}$ (FWHM), while the NVSS
used the most compact D array with a flux density limit of $\sim 2.5$~mJy/beam
($sim 0.5$~mJy/beam rms) and an angular resolution of $\sim 45^{\prime\prime}$.

Given the relatively low sensitivity, and the similar ($u,v$) sampling with
our 6~cm observations, we expect to detect all the ZGS sources in our
maps
(see also Becker et al.~\cite{Bea94}). On the other hand, due to 
the much higher sensitivity of the NVSS and its ability to detect
extended structures, many of the fainter 20 cm sources 
with non-thermal spectral indexes and/or sizes greater than 10$^{\prime\prime}$
will not
be detectable in our observations. In Fig.~\ref{fpos}~b) we show the 
positions of all the ZGS (11 -- pluses) overlaid on the contour plot of
the NVSS image of our survey region.

In Table~\ref{tass} the results of the correlation between our catalogue and
the 20~cm surveys is presented. The relevant parameters (names, positions,
flux densities and sizes) of the 20~cm sources are from the published catalogues
(Zoonematkermani et al.~\cite{Zea90} for the ZGS and the deconvolved data
from the fits catalogue available on the World Wide Web at
{\tt http://www.nrao.edu} in October~1998 for the NVSS).
The matching criterion used is positional coincidence: ZGS sources 
are considered to be associated with our sources if the positional difference
is less than half a beamwidth for point sources, or if the source position falls
within the boundary of one of our extended sources; NVSS sources are considered
to be associated if one of our point source falls inside of,
or if the boundaries of one of our extended sources
overlap with, the deconvolved size of the 20~cm source.
As expected, all the ZGS sources in our surveyed field do have a 
counterpart at 6~cm. In one case (source \#32 in our list),
we considered two ZGS sources as being part of the same (extended) 6~cm source.
In Table~\ref{tass}, columns~1 and~2 report the numbers and names of our
sources from Table~\ref{tsrc}, columns~3 to 6 the names, peak and 
integrated flux densities, and sizes of the ZGS sources, columns~7 to 10 the names,
integrated flux densities and deconvolved sizes of the NVSS sources,
and column~11 the IRAS sources names (see Sect. 3.4).

In general, given the higher sensitivity of the NVSS and its ability to 
detect extended sources that might be resolved out in the ZGS, we expect that 
all the ZGS sources in our field should be detected in the NVSS as well.
The only possible exception is that of very compact high surface brightness 
sources close or inside large low surface brightness sources with high
integrated flux. There are 3 ZGS sources without an NVSS counterpart,
one (045.129$+$0.131, associated to our \#17) is indeed inside the bright 
complex shown in Fig.~\ref{exfig2}, and thus may be missing from the NVSS 
catalogue due to confusion. Similarly, the one associated with our \#19 could
be undetected in the NVSS due to its proximity to the extended source \#33.
Both \#17 and \#19 have thermal spectral indexes (see below and
Table~\ref{tspecind}) and we do not expect them to be variable at
20~cm. On the other hand, the ZGS source associated with \#29 should have been
detected in the NVSS, thus for this source, given also its non-thermal
spectral index, the only viable explanation for the NVSS non-detection
is variability at 20~cm.

Finally, there is a very bright ($\sim$280~mJy), unresolved, NVSS source which
is undetected in the ZGS and in our survey. This source
(clearly visible in Fig.~\ref{fpos}~b) at $l\sim 45.35$, $b\sim -$0.22)
is the high energy source G1915$+$105 (Mirabel \& Rodr\'{\i}guez~\cite{MR94}).
At radio wavelengths is known to be highly variable, with flux densities
at 1.4~GHz that can be as high as $\sim 1$~Jy at the peak of radio
bursts and below the mJy level during quiescence (Rodr\'{\i}guez
\& Mirabel~\cite{RM99}).

\begin{table*}
\caption[]{\label{tass}Associated ZGS, NVSS and IRAS--PSC2 sources}
\begin{tabular}{rllrrrlrrrl}
\hline
 && \multicolumn{4}{c}{ -------------------------- ZGS -------------------------- } & \multicolumn{4}{c}{ --------------------- NVSS$^a$ --------------------- } & IRAS\\
\#&Name& Name & \multicolumn{1}{c}{F$_p$} & \multicolumn{1}{c}{F$_i$} & \multicolumn{1}{c}{d} & Name$^a$ &\multicolumn{1}{c}{F$_i$} & \multicolumn{1}{c}{Size$^b$} & \multicolumn{1}{c}{p.a.$^b$} & Name \\
&&&\multicolumn{1}{c}{(mJy/b)}&\multicolumn{1}{c}{(mJy)}&\multicolumn{1}{c}{($^{\prime\prime}$)}&(NVSS J)&\multicolumn{1}{c}{(mJy)}&\multicolumn{1}{c}{($^{\prime\prime}\times^{\prime\prime}$)}&\multicolumn{1}{c}{($^\circ$)}&\\
\hline
5&G044.792$+$0.380&               &    &    &    &191156$+$104256&2.8&129$\times$55&0& \\
9&G044.965$+$0.284&               &    &    &    &191236$+$104930&3.3&58$\times$48&0& \\
14&G045.101$+$0.122&045.101$+$0.121&  41&  49& 2.2&191327$+$105217&58.2&64$\times$21&22& \\
15&G045.123$+$0.132&045.123$+$0.132& 287&1468&10.4&191327$+$105338&1540.4&21$\times$15&58& 19111$+$1048\\
17&G045.130$+$0.131&045.129$+$0.131&  22&  93& 9.1& & & & &\\
19&G045.466$+$0.045&045.466$+$0.046&  16&  21& 2.6& & & & &\\
25&G044.996$-$0.446&044.995$-$0.445&  21&  22& 0.0&191518$+$103042&20.4&22$\times$29&83& \\
27&G044.967$-$0.544&044.967$-$0.543&  21&  23& 1.7&191536$+$102630&14.2&32$\times$27&0& \\
29&G045.007$-$0.614&045.007$-$0.614&  12&  10& 0.0& & & & &\\
30&G045.066$+$0.138&               &    &    &    &191320$+$105054&401.9&31$\times$19&$-$83&19110$+$1045\\
31&G045.134$+$0.145&045.134$+$0.143&  48&2245&34.8&191326$+$105422&2445.3&46$\times$42&53&\\
32&G045.479$+$0.130&045.477$+$0.130&  97&1222&17.0&191408$+$111229&1672.6&33$\times$14&$-$33&19117$+$1107\\
&                &045.480$+$0.136&  62& 653&15.2& & & & & \\
33&G045.455$+$0.059&045.454$+$0.060& 167&2207&17.5&191421$+$110913&4771.5&41$\times$36&$-$20&19120$+$1103\\
34&G045.190$-$0.439&               &    &    &    &191539$+$104123&50.8&20$\times$17&$-$41&19132$+$1035\\
\hline
\end{tabular}
\vskip 0.3cm
$^a$) In this table, the ``NVSS J'' prefixes have been omitted from the
names of the NVSS sources.\\
$^b$) Deconvolved major, minor axes and position angle
(see Cotton et al.~\cite{Cea98}).\\
\end{table*}

In Table~\ref{tspecind}, columns~2 to 6, we report the radio continuum
spectral indexes ($\alpha$, defined as $F_\nu\sim\nu^\alpha$)
as calculated from our integrated flux densities and the ZGS and
NVSS integrated flux densities. It should be noted that all extended sources are 
probably partially resolved out in the higher resolution surveys,
particularly in our 3.6~cm images, and thus some of the measured spectral
indexes are probably lower limits due to the missing flux at high frequency.

\begin{figure}
\centerline{\psfig{figure=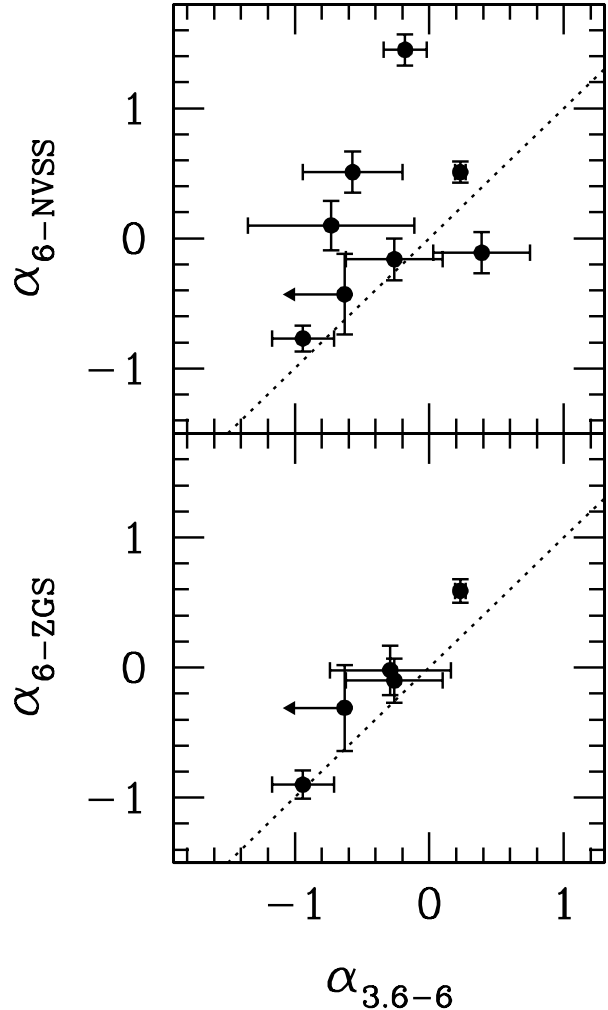,width=8.5cm}}
\caption[]{\label{fspec}Comparison between the high and low frequency 
spectral indexes. Top panel: sources in our survey detected at 20~cm
in the NVSS. Bottom panel: sources detected in the ZGS. In both panels
the dotted line represent equal spectral indexes.}
\end{figure}

In Fig.~\ref{fspec} we compare the high frequency spectral indexes 
(those calculated between 3.6 and 6~cm) with the low frequency ones
(calculated between 6 and 20~cm), only the sources inside the area 
observed both at 3.6 and 6~cm have been considered. A 10\% error has
been assumed for all ZGS and NVSS integrated flux densities (this may be a slight
underestimate of the true error for the faintest sources in these surveys).
In the upper panel we show the comparison for sources detected in the
NVSS and in the lower panel that for sources detected in the ZGS.
We find very good agreement between the high frequency and the low
frequency spectral indexes for ZGS sources. This is probably due to the
matched beams of the observations. In contrast, for NVSS sources,
the spread between low and high frequency spectral indexes is
much greater. There are two possible explanations for this: 1) the increased
sensitivity to extended structures of the NVSS and 2) the greater sensitivity
of the NVSS with respect to the ZGS. The increased sensitivity allows
for the detection in the 
NVSS of some thermal sources that are optically thick at 
low frequency and become optically thin at high frequency (this
is probably the case for \#9 and \#34).

\begin{table*}
\caption[]{\label{tspecind}Radio continuum spectral indexes and IRAS
fluxes for the detected sources}
\begin{tabular}{lrrrrrrrrrl}
\hline
\#
&$\alpha_{3.6-6}$&$\alpha_{\rm 6-ZGS}$&$\alpha_{\rm 3.6-ZGS}$&$\alpha_{\rm 6-NVSS}$&$\alpha_{\rm 3.6-NVSS}$&F$_{12\mu\rm m}$&F$_{25\mu\rm m}$&F$_{60\mu\rm m}$&F$_{100\mu\rm m}$& ID\\
&&&&&&(Jy)&(Jy)&(Jy)&(Jy)&\\
\hline
01 & $-$1.29$\pm$0.90 &  --            & --             &  --            & --              &&&&&\\
02 & $-$0.13$\pm$1.10 &  --            & --             &  --            & --              &&&&& Cand~HII\\
03 & $>+$1.01       &  --            & --             &  --            & --              &&&&&Cand~HII\\
04 & $<-$1.50       &  --            & --             &  --            & --              &&&&\\
05 & $-$0.73$\pm$0.62 &  --            & --             & $+$0.10$\pm$0.19 & $-$0.16$\pm$0.17  &&&&\\
06 & $-$1.07$\pm$1.38 &  --            & --             &  --            & --              &&&&\\
07 & $<-$0.06       &  --            & --             &  --            & --              &&&&\\
08 & $+$0.93$\pm$0.90 &  --            & --             &  --            & --              &&&&&Cand~HII\\
09 & $-$0.18$\pm$0.16 &  --            & --             & $+$1.45$\pm$0.12 & $+$0.95$\pm$0.07  &&&&&Cand~HII\\
10 & $-$0.77$\pm$0.76 &  --            & --             &  --            & --              &&&&\\
11$^a$ & $+$0.86$\pm$0.30$^a$ &  --            & --             &  --            & --              &&&&&Cand~HII\\
12$^a$ & $+$0.86$\pm$0.30$^a$ &  --            & --             &  --            & --              &&&&&Cand~HII\\
13 & --$^b$         &  --            & --             &  --            & --              &&&&\\
14 & --$^b$         & $-$0.31$\pm$0.33 & --$^b$           & $-$0.43$\pm$0.31 & --$^b$            &&&&\\
15 & $+$0.23$\pm$0.04 & $+$0.59$\pm$0.09 & $+$0.47$\pm$0.07 & $+$0.51$\pm$0.08 & $+$0.42$\pm$0.06  &250&1400&5900&7500&HII\\
16 & --$^b$         &  --            & --             &  --            & --              &&&&\\
17 & $-$0.29$\pm$0.45 & $-$0.02$\pm$0.19 & $-$0.11$\pm$0.13   &  --            & --              &&&&&Cand~HII\\
18$^c$ &  --            &  --            & --             &  --            & --              &&&&&HII$^c$\\
19 &  --            &  --            & $+$0.94$\pm$0.11 &  --            & --              &&&&&Cand~HII\\
20 & $<-$0.46       &  --            & --             &  --            & --              &&&&\\
21 & $<-$1.21       &  --            & --             &  --            & --              &&&&\\
22 & $-$0.06$\pm$0.67 &  --            & --             &  --            & --              &&&&&Cand~HII\\
23 & $<-$2.33       &  --            & --             &  --            & --              &&&&\\
24 & $<-$0.62       &  --            & --             &  --            & --              &&&&\\
25 & $-$0.94$\pm$0.23 & $-$0.90$\pm$0.11 & $-$0.91$\pm$0.12 & $-$0.77$\pm$0.10 & $-$0.83$\pm$0.11  &&&&\\
26 & $<-$2.38       &  --            & --             &  --            & --              &&&&\\
27 &  --            & $-$1.03$\pm$0.13 & --             & $-$0.57$\pm$0.12 & --              &&&&\\
28 &  --            &  --            & --             &  --            & --              &&&&\\
29 &  --            & $-$1.09$\pm$0.17 & --             &  --            & --              &&&&&Variable?\\
30 & $+$0.39$\pm$0.36 &  --            & --             & $-$0.11$\pm$0.16 & $+$0.04$\pm$0.11  &58&490&$<$5900&$<$7500&HII\\
31 & $-$0.26$\pm$0.36 & $-$0.10$\pm$0.17 & $-$0.15$\pm$0.12 & $-$0.16$\pm$0.16 & $-$0.19$\pm$0.11  &&&&&HII\\
32 &  --            &  --            & $-$0.13$\pm$0.12 &  --            & $-$0.06$\pm$0.11  &37&303&2600&$<$7900&HII\\
33$^c$ &  --            &  --            & $+$0.42$\pm$0.10 &  --            & $-$0.03$\pm$0.10  &79&640&5300&7900&HII$^c$\\
34 & $-$0.57$\pm$0.37 &  --            & --             & $+$0.51$\pm$0.16 & $+$0.18$\pm$0.11  &6.9&34&280&490&HII\\
\hline
\end{tabular}
\vskip 0.3cm
$^a$) Sources \#11 and \#12 are blended together at 6~cm, the separation 
of the two contribution to the integrated flux is very uncertain, thus we 
calculated the spectral index using the integrated flux of both
components together\\
$^b$) For these sources, due to the confusion and noise in the 8.4~GHz map
it is difficult to obtain a reliable estimate of the upper limit on the
spectral index\\
$^c$) Source \#18 is inside the extended source \#33. The total integrated
flux density (4.5~Jy) has been used to determine the spectral indexes
(reported only for \#33).
\end{table*}

\subsubsection{NVSS sources undetected at high frequency}

Most of the NVSS sources in our field (48) are not detected at 6 and/or 3.6~cm.
We believe that in most cases the negative spectral index, rather
than the different ($u,v$) coverage between the observations, is the
main reason for the non-detection at high frequency. 
The most plausible explanation is that a large fraction of these NVSS sources
are extragalactic objects, with a possible contamination from faint planetary
nebulae. 

\begin{figure}
\centerline{\psfig{figure=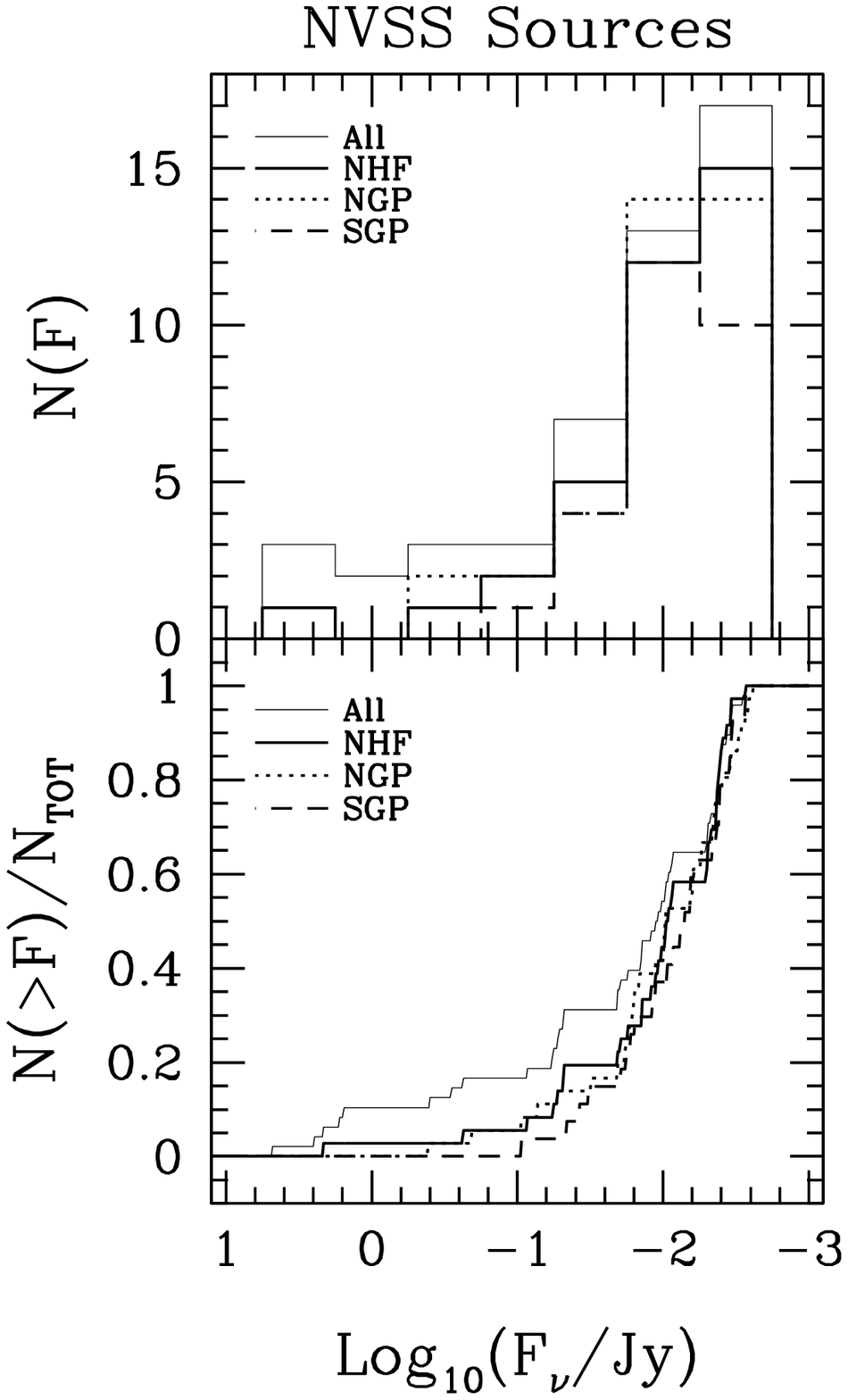,width=8.5cm}}
\caption[]{\label{fnvsslf}Top panel: differential luminosity functions
of all (All; thin continuous line) and not detected at high frequency
(NHF; thick continuous line) NVSS sources inside our field, and
of NVSS sources from two 0.652~sq.deg. areas close to the northern
(NGP; dotted line) and southern (SGP; dashed line) galactic poles. Bottom panel:
cumulative luminosity functions for the same sources shown in the upper
panel.}
\end{figure}

To check whether the 20~cm flux distribution and source count for the NVSS
sources not detected at high frequency are consistent with the population
of extragalactic radio sources, we extracted from the NVSS the sources 
in two areas toward the galactic poles, each of the two
with the same extent of our surveyed region. The number of sources 
extracted toward the northern and southern galactic poles are 36 and 27,
respectively, these numbers compare relatively well with the
37 NVSS sources without high frequency counterpart in our field. 
As additional check, in Figure~\ref{fnvsslf},
we show the differential and cumulative luminosity functions for the sources
in our field and those in the areas toward the galactic poles. The 
luminosity function of all the sources in our field (thin line) show an excess 
of bright sources with respect to the galactic poles, this excess 
disappears if we plot only the sources without a high frequency 
counterpart (thick line). This effect is more clear in the cumulative
luminosity function plot (Fig.~\ref{fnvsslf}, lower panel).
More quantitatively, the Kolmogorov-Smirnov test on the cumulative luminosity
functions gives a probability lower than 40\% that the NVSS sources in
the Galactic poles samples and those in our field are drawn from the same
distribution. This probability rises above 80\% if we remove from our
sample the sources detected at high frequency and the well known galactic
high energy source G1915$+$105.

\subsection{Effelsberg 5~GHz survey}
\label{salt}

As mentioned in Sec.~\ref{sintro}, our surveyed region has been covered
by the Altenhoff et al.~(\cite{Aea78}) 5~GHz (6~cm) single dish survey. 
The names and peak flux densities of the seven single dish sources inside or
partially within our survey boundaries are listed in Table~\ref{talt}.
In the same table, for each source, we report the integrated flux densities of
our VLA 6~cm sources within the Effelsberg beam (2.6$^\prime$).

\begin{table}
\caption[]{\label{talt}Comparison between single dish and VLA 5~GHz sources.}
\begin{tabular}{lrrl}
\hline
\multicolumn{2}{c}{Effelsberg}&VLA&\\
Name & F$_p$ & F$_i$ & Sources ID from\\
     & (Jy)  & (Jy)  & Table~\ref{tsrc} \\
\hline
44.786$-$0.490 & 0.2 & --    & Not detected, high rms\\
45.066$-$0.135 & 0.7 & 0.62  & 11, 12, and 30\\
45.125$+$0.136 & 5.8 & 5.25  & 13--17, and 31\\
45.202$-$0.411 & 0.2 & 0.096  & 34\\
45.341$-$0.370 & 0.2 & 0.002 & 26$^a$ \\
45.451$+$0.060 & 6.4 & --    & Not mapped at 6~cm\\
45.475$+$0.130 & 2.1 & --    & Not mapped at 6~cm\\
\hline
\end{tabular}
\vskip 0.3cm
$^a$) This source is known to be variable (e.g. Harmon et al.~\cite{Hea97}).
\end{table}

For one of the single dish sources (44.786$-$0.490) the peak is 
outside our survey area. We do not detect this source at either 6 or 3.6~cm,
probably because it is resolved out in our interferometric observations.
The last two sources in Table~\ref{talt}
are in the region covered only at 3.6~cm, they have been detected at this
wavelength and correspond to sources [18+19+33] and 32 in Table~\ref{tsrc}.
The other four sources have been detected in our 6~cm observations,
and our integrated flux densities are in reasonable agreement with the single
dish ones, except for 45.341$-$0.370 (our source 26) which is known
to be a highly variable source (see e.g. Harmon et al.~\cite{Hea97}).
Somewhat surprisingly, in our VLA 6~cm images we recover almost all the
single dish flux for the extended complexes 45.066$-$0.135 and
45.125$+$0.136, while about half of the single dish flux is 
recovered for 45.202$-$0.411.

\subsection{IRAS Point Sources Catalogue}

To search for far infrared (FIR) counterparts to our detected radio sources,
we extracted from the IRAS-PSC2 (Beichman et al.~\cite{Bea88}) catalogue
all the sources inside our survey area.
In Figure~\ref{fpos}~c) we show the positions of all (43)  IRAS point 
sources inside the observed field. We could find an IRAS counterpart 
within 100$^{\prime\prime}$ only for 5 of our 3.6 and/or 6~cm sources.
In all five cases, the IRAS error ellipse contains the radio continuum
source or overlaps with the boundaries of the extended radio sources.
In fact, in all five cases the distance from the peak of the radio continuum
source and the nominal IRAS position is less than 30$^{\prime\prime}$.
The FIR fluxes of these five sources are reported in Table~\ref{tspecind},
columns~7 to 10.

\begin{figure*}
\centerline{\psfig{figure=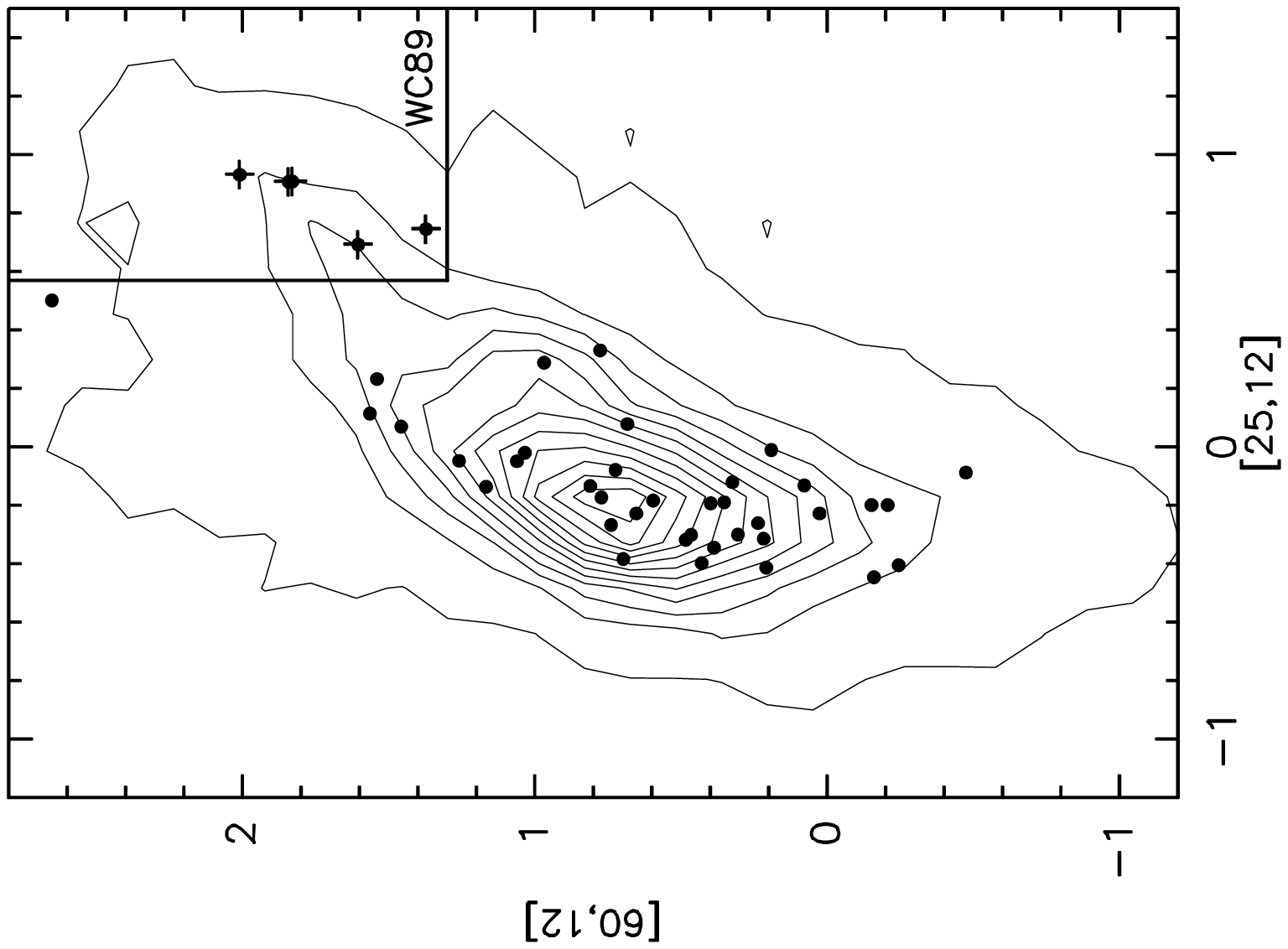,width=5.5cm,angle=-90}
            \hskip 0.4cm
            \psfig{figure=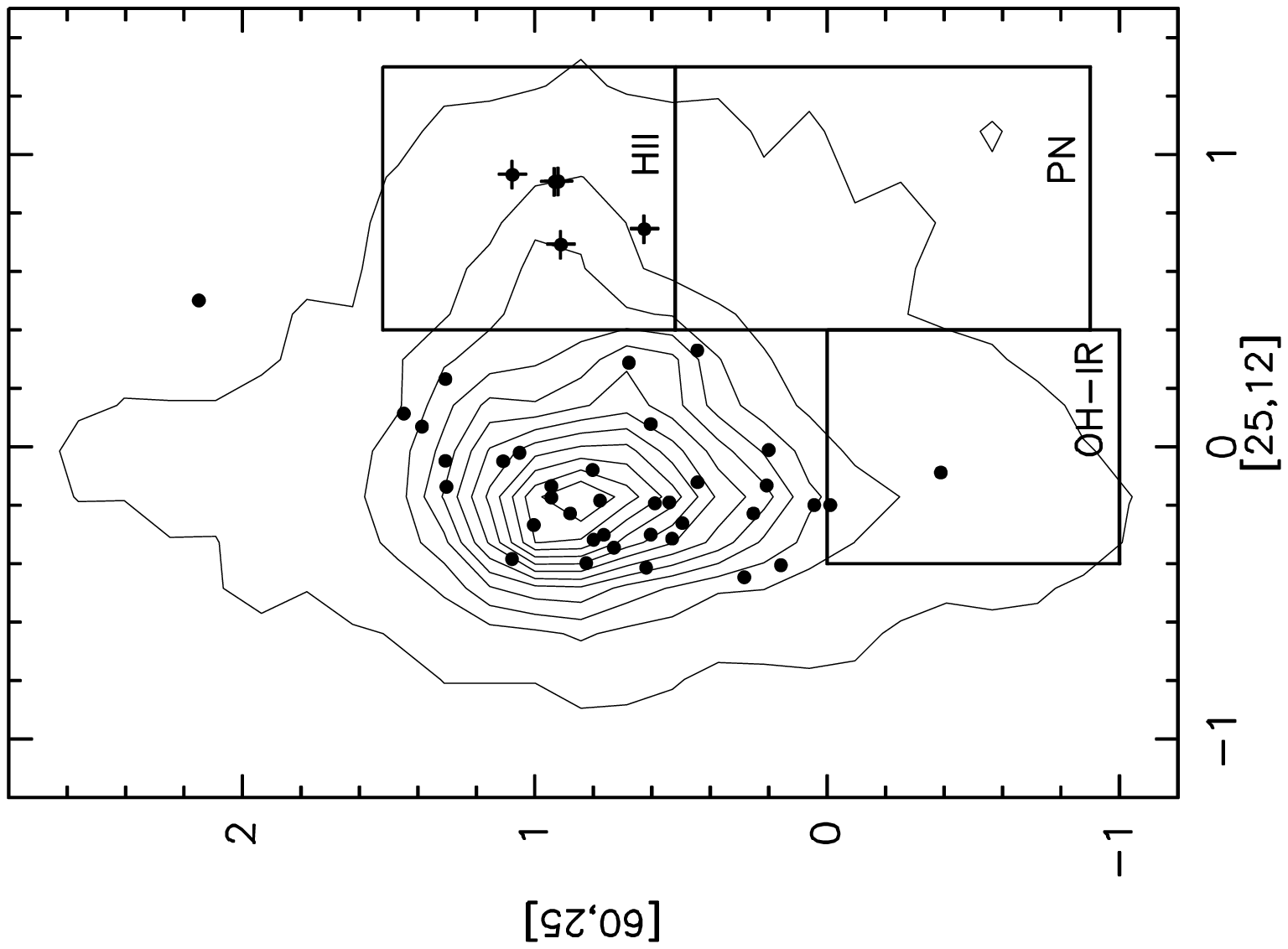,width=5.5cm,angle=-90}
            \hskip 0.4cm
            \psfig{figure=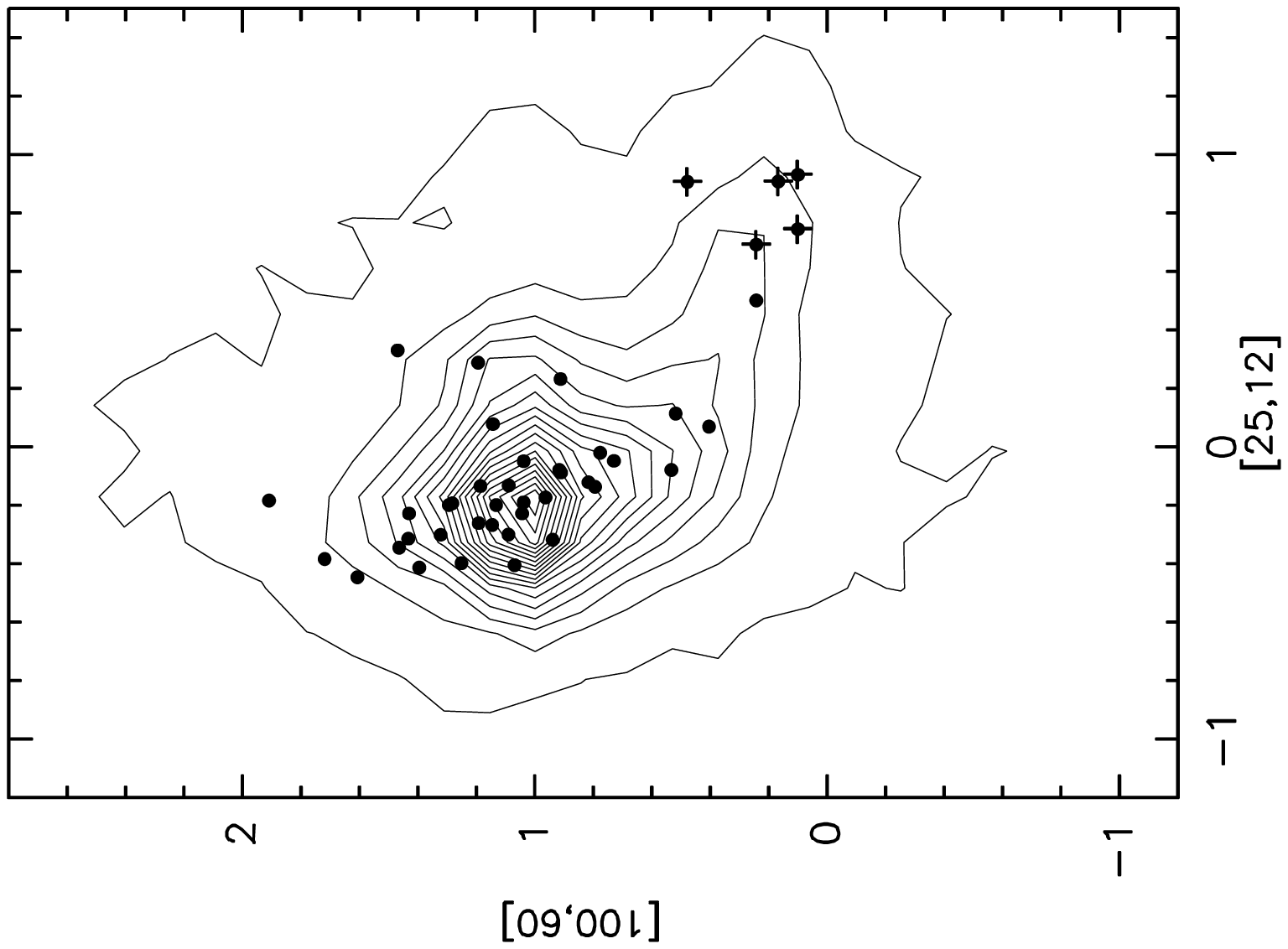,width=5.5cm,angle=-90}}
\caption[]{\label{firascc}
In each diagram:
[$\lambda_i$,$\lambda_j$]$\equiv\log_{10}(F_{\lambda_i}/F_{\lambda_j})$;
the contour plots represent
the normalized surface density of IRAS-PSC2 sources in the region
$|l|\le 90^\circ$, $|b|\le0^\circ\!\!.65$.
Black filled circles show the colors of the 43 sources in our surveyed region
(one source detected only at 100~$\mu$m is not present in the first two
plots). Many of the sources only have upper limits at one or more IRAS bands,
the colors for these sources are either upper, lower limits, or are 
undetermined. We have not marked these sources with special symbols
as we have not corrected the color-color surface 
density contours for upper limits. The five IRAS sources with a radio continuum counterpart
are marked with plus symbols.
}
\end{figure*}

The study of the IRAS color-color diagram is a powerful tool to 
investigate the nature of the FIR sources. Different types of 
objects tend to populate different parts of the color-color planes.
In Fig.~\ref{firascc} we show three of the color-color diagrams
that can be constructed using the four IRAS fluxes, and that have
been shown to be able to separate different types of galactic sources
(e.g. Eder, Lewis \& Terzian~\cite{ELT88};
Pottasch et al.~\cite{Pea88}; WC89
White, Becker \& Helfand~\cite{WBH91}).
In each diagram the contour plots represent
the normalized surface density of the colors
([$\lambda_i$,$\lambda_j$]$\equiv
log_{10}(F_{\lambda_i}/F_{\lambda_j})$) of IRAS-PSC2 sources 
within the inner galactic plane, defined as:
$|l|\le 90^\circ$, $|b|\le0^\circ\!\!.65$.

We note that the 43 IRAS sources in our field tend to populate the
color-color planes in the same fashion as the entire inner galactic
plane sample (contour plots), which, of course, is what we expected.
It is remarkable, however, that all, and only, the IRAS sources detected
in radio continuum (marked with plus symbols in the figure) lie in
well-defined, low-density parts of the planes.
This is the part of the color-color planes where ultra compact HII (UCHII)
regions are expected to be found (WC89; Wood \& Churchwell~\cite{WC89b};
Kurtz, Churchwell \& Wood~\cite{KCW94}; White et al.~\cite{WBH91};
Becker et al.~\cite{Bea94}).

\section{Discussion}
\label{sdis}
\subsection{Classification of radio sources}

We shall now classify the sources detected in our survey and reported 
in Table 1 according to morphology,
spectral index and coincidence with IRAS sources.

Five complexes of sources have all the specifications for being 
classified as thermal galactic HII regions. They include all the extended
sources plus some additional small diameter source in the same area, 
more precisely [11 + 12 + 30], [13 + 15 + 16 + 17 + 31],~[34],~[18 + 33] 
and [32] (numbers are from Table~\ref{tsrc}).
All these complexes coincide with corresponding sources in the Altenhoff et
al.~\cite{Aea78} survey (see Sect.~\ref{salt}) and are now resolved
in much more detail.
Morphologically they show the classical aspect of a cluster of HII
regions, of which G9.62+0.19 is a typical example (Testi et al.~\cite{TFPR98};
\cite{THKR99}), i.e. several sources of different angular sizes
(from unresolved to several tens of arcsec) are clustered in the same
area.  The continuum sources may represent independent UCHII
regions born in the same star forming complex but presently observed in
different evolutionary phases with the unresolved sources being the youngest
and the more extended sources more evolved. 

Six of the small diameter sources (2, 3, 8, 9, 19, 22) can be
classified as ``candidate HII region'' according to their spectral
index. No IRAS source is associated with them,  but their radio flux
is rather low. Non detection at far infrared wavelengths could
be either due to the intrinsic weakness of some of these sources or,
most probably, due to the incompleteness of the IRAS-PSC in the 
galactic plane (see also Becker et al.~\cite{Bea94}).

The remaining 15 sources (4, 5, 6, 7, 10, 14, 20, 21, 23, 24,
25, 26, 27, 28 and 29) can be classified  from the spectral
index as non thermal (probably extragalactic) sources. 
Only five of these have been detected at 20 cm. These have in general
greater integrated flux densities at 6 cm than those not detected at 20 cm
(the mean 6~cm flux densities of the two groups are 10 and 2 mJy,
respectively), so that the effect can be simply explained as due to 
our higher sensitivity at 6 cm. All 15 sources have been detected at
6~cm and 4 of them at 3.6~cm as well. Given the area observed at 6~cm
(0.620~sq.~deg.) and that observed at 3.6~cm (0.525~sq.deg.),
the number of extragalactic sources above the 1~mJy threshold,
which we can assume as a mean detection limit for our survey,
can be estimated from deep VLA surveys.
Following Fomalont et al.~(\cite{Fea91})
at 6~cm we expect 15 extragalactic sources above our 1~mJy threshold, while
at 3.6~cm the number is reduced to 9 sources for the same threshold
(Windhorst et al.~\cite{Wea93}). Given the small number statistics,
these numbers are in relatively good agreement with the source counts
in our surveyed area.

Becker et al.~(\cite{Bea94}) estimated a total of $\sim 100$ planetary
nebulae (PNs) down to a flux limit of $\sim 2.5$~mJy in their 6~cm survey 
of 50~sq.deg. of the inner galactic plane. This number correspond
to less than 2 PNs expected down to the same flux level in our 6~cm
survey region. Thus the contamination from PNs in our source lists
should be very small.

\subsection{IRAS ``UCHII-type'' sources}

In Sect. 3.4 it was pointed out that all the IRAS sources with a corresponding
radio counterpart in our survey (5 out of 43) satisfy the color-color
criteria to be classified as UCHII regions (WC89). 
However, with the possible exception of the double source G045.070$+$0.132
and G045.072$+$0.132 (11 and 12), none of the radio sources within
100$^{\prime\prime}$ from the IRAS-PSC position can be 
classified as {\it bona fide} UCHII region using the usual definition (Wood \&
Churchwell~\cite{WC89b}; Kurtz et al.~\cite{KPPIV99}). The radio continuum
sources are extended (non homogeneous) HII regions, with emission peaks
inside them that may appear as UCHII regions when observerved with an extended
VLA configuration.
A tipical example could be G045.455$+$0.060 which appears as a compact
source inside the extended HII region G045.455$+$0.059 (see
Figure~\ref{exfig3}), this source has the appearence of an UCHII region
in the Wood \& Churchwell~(\cite{WC89b}) survey (their source G45.45$+$0.06).
The VLA high frequency and high resolution surveys of 
IRAS selected UCHII candidates are all biased to the detection of
only the most compact and dense ionized gas, due to the spatial filtering
of the interferometer, and are unable to detect the extended components.
Our results agree with those of Kurtz et al.~(\cite{K99}) and show that, when 
observed with sufficient sensitivity to extended structures, most, if not
all, the IRAS selected UCHII candidates do have extended radio components.
This implies that samples of IRAS-PSC sources selected with the WC89 criteria
are contaminated by a {\it substantial} number of {\it older} more extended HII 
regions (see also Codella, Felli \& Natale~\cite{CFN94};
Ramesh \& Sridharan~\cite{RS97}; Kurtz et al.~\cite{K99}).
The number of UCHII regions estimated from the color selected
IRAS-PSC counts may be, consequently, overestimated by a large factor.
If most IRAS-WC89 sources are indeed associated with extended 
HII rather than UCHII regions, the lifetime of the IRAS-WC89 color phase of 
UCHII/HII regions may be much longer than estimated from the early
high resolution radio surveys. Consequently, the estimated lifetime of the 
UCHII phase for O-type young stars is probably much shorter that previously
thought (see also
Ramesh \& Sridharan~\cite{RS97}). Additionally, we find 6 UCHII candidates 
in our radio continuum survey without an associated IRAS source.
As discussed by Becker et al.~(\cite{Bea94}), this is probably due
to the confusion limit of the PSC on the galactic plane, and the 
generally lower radio luminosity of these sources. However, 
we note that in our field {\it only} unresolved thermal radio sources are
not present in the IRAS-PSC, while {\it all} resolved HII 
regions are detected in the far-infrared. Incidentally, we note that all the 
compact thermal radio sources in our survey not associated with IRAS
PSC sources are fainter at centimeter wavelengths than those detected
in the far infrared, and thus they may be associated with 
stars of type later than O. However, without knowing the distances it is
impossible to draw a final conclusion.

In our surveyed region, the percentage of IRAS sources satisfying
WC89 color criteria is (5$/$43$\sim$12\%). This is consistent
with the percentage found accross the entire inner galactic plane
($|l|\le 90^\circ$, $|b|\le 0^\circ\!.6$, $\sim 8$\%). The fraction of 
WC89 sources in the IRAS-PSC database drops to much lower values
outside the inner galactic plane (WC89). 

\subsection{Continuum emission from the H$_2$O maser}

During an incomplete low spatial resolution (2$^\prime$) single dish survey of
the $l=+45^{\circ}$ field in the H$_2$O 22~GHz maser line, a new maser
was detected. The masing gas is probably coincident with a 15 $\mu$m
source (F$_{15{\mu}\rm m}$ = 370 mJy) located at
$\alpha(2000)=19^{\rm h}12^{\rm m}46^{\rm s}$ $\delta(2000)=10^\circ45^\prime30^{\prime\prime}$,
and was interpreted as a candidate young stellar object (Testi et
al.~\cite{Tea97}). Therefore, it was interesting to see if any
radio continuum emission from an associated UC HII region could be
detected.

From a careful inspection of the area around the maser, no radio
continuum emission was seen above the (local) 3$\sigma$ level
(0.6~mJy/beam at 3.6~cm and 1.2~mJy/beam at 6~cm).
With the young stellar object hypothesis in mind, there are two
possible explanations: 1) the putative UCHII region is intrinsically too 
weak to be detected or absent because the eventual exciting star is of
late spectral type; or  2) there is an UCHII region, but it is in such an 
early evolutionary phase that it is  
optically thick even at 3.6 cm. The lack of radio continuum emission
close to H$_2$O masers in high luminosity star forming regions
has been amply demonstrated by a survey of a large number of maser
in the radio continuum, which showed that many maser associated
with high luminosity sources do not have any close-by radio continuum source
(Tofani et al.~\cite{TFTH95}). Subsequent molecular observations
of the masers without continuum emission has indeed confirmed that
these are associated with very young star forming regions since in all
cases a hot molecular  core was found at the same position (Cesaroni et
al.~\cite{CFW99}). 

To settle the nature of the new maser - 15$\mu$m source, molecular
observations in high density tracers are needed, as well as an estimate
of its luminosity.

\section{Conclusions}
\label{scon}

The unbiased radio continuum survey of the ISOGAL field at 
$l=+45^{\circ}$ has resolved the structure of five thermal
extended complexes and discovered 21 additional small diameter sources,
six of which are candidate HII regions.

Comparison with the IRAS PSC shows that all 5 of the extended thermal
sources have corresponding FIR emission and that the colors of these
sources satisfy the WC89 color criteria for UCHII. Our sources, however,
are {\it not} UCHII regions, but are more evolved extended HII regions.
This result is consistent with the results of earlier single dish 
surveys (Codella et al.~\cite{CFN94}) and of a recent survey for 
extended emission around IRAS-selected UCHII regions(Kurtz et al.~\cite{K99}).

We conclude that UCHII counts based on IRAS selected samples are overestimated
by a large factor, consequently the estimated lifetime of the
UCHII phase may be substantially reduced, removing the so-called
lifetime problem for UCHII regions.

The percentage of IRAS sources associated with  HII regions
is $\sim$10\% in our field, which seems  to be a general
property of IRAS sources in the galactic plane.

\begin{acknowledgements}
Support from CNR-NATO Advanced Fellowship program and 
from NASA's {\it Origins of Solar Systems} program (through grant NAGW--4030)
is gratefully acknowledged.
\end{acknowledgements}

\appendix
\section{Maps of all the detected radio sources}
\label{acp}

\begin{figure*}
\centerline{\psfig{figure=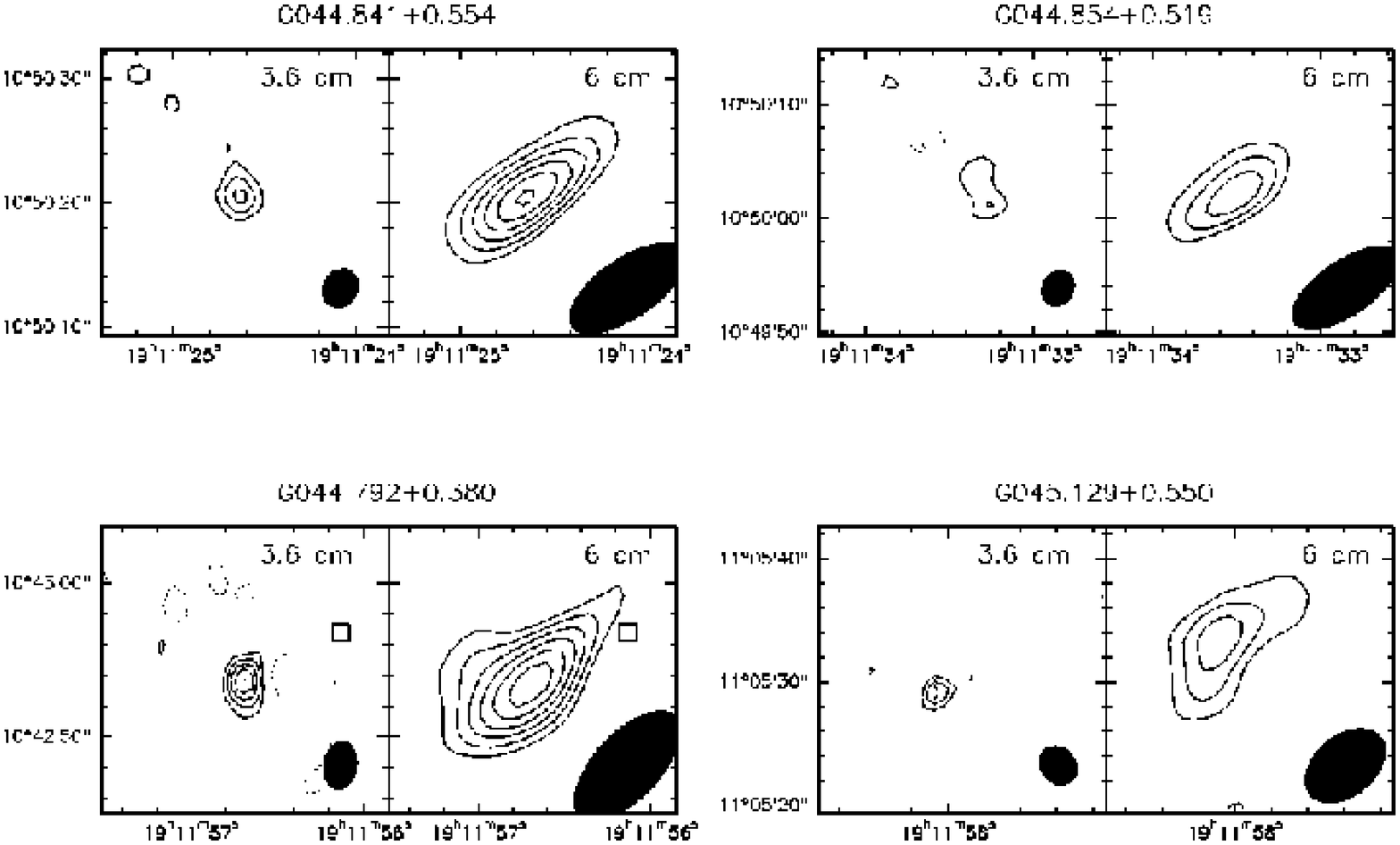,width=14.2cm,angle=-90}}
\vskip 0.9cm
\centerline{\psfig{figure=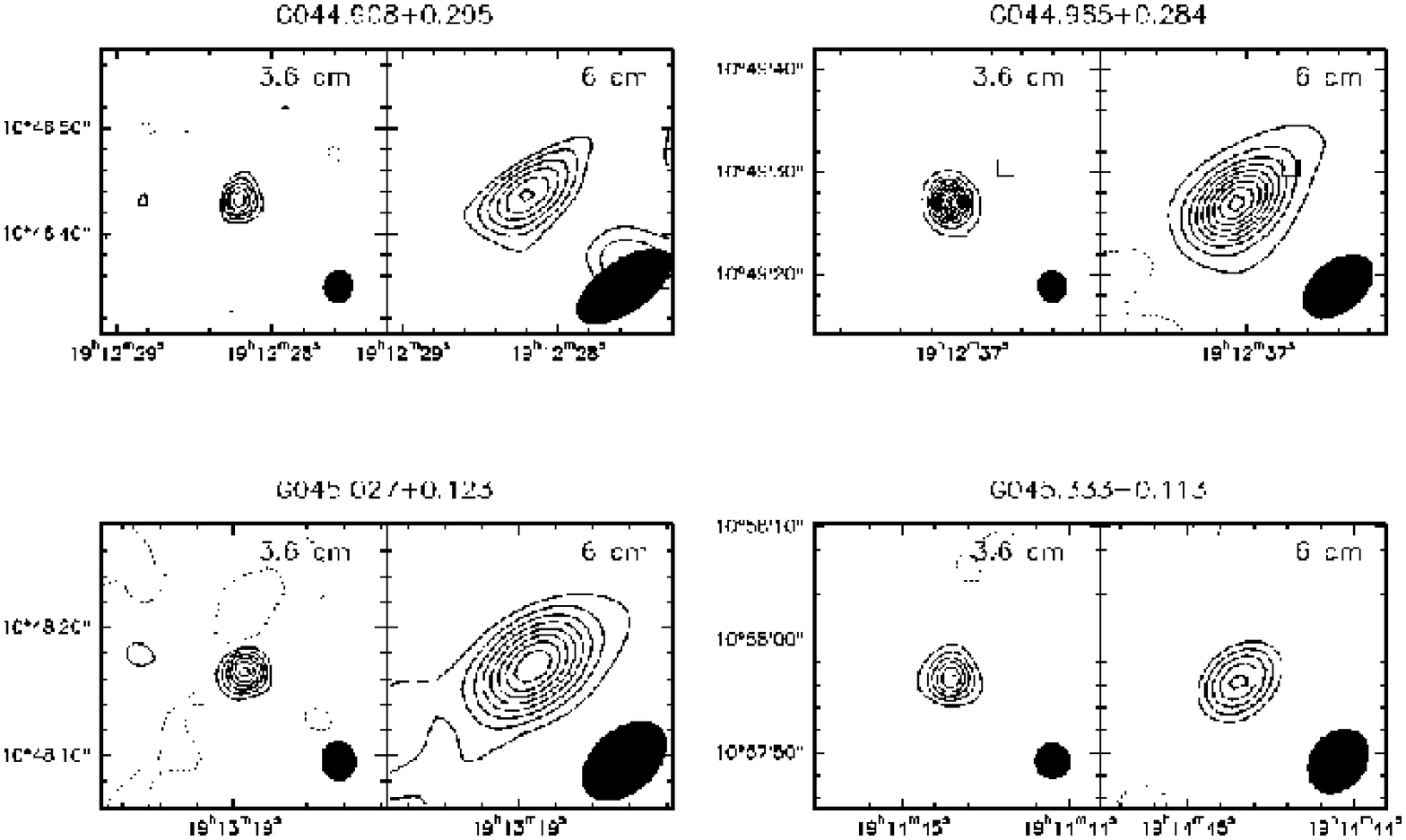,width=14.2cm,angle=-90}}
\vskip 0.9cm
\centerline{\psfig{figure=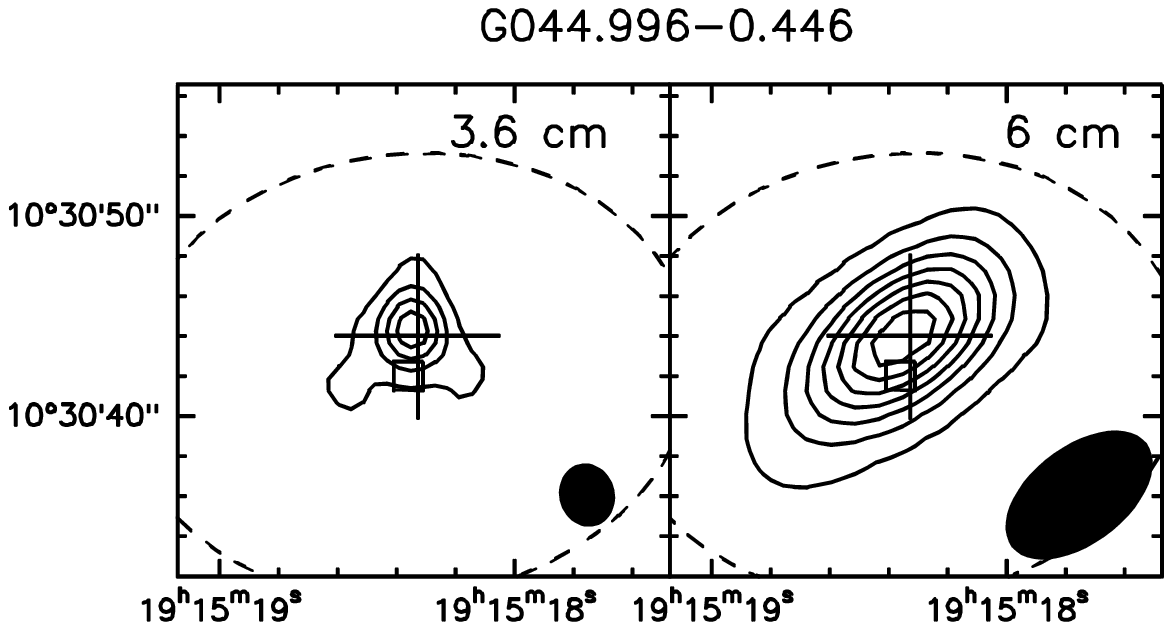,width=14.2cm,angle=-90}}
\vskip 0.4cm
\caption[]{\label{pcfig1}3.6 and 6~cm maps of the unresolved sources
detected at both frequencies; crosses mark the peak positions
of 20~cm sources from Zoonematkermani et al.~(\cite{Zea90});
open sqares mark the positions of the peaks of NVSS sources (Condon et
al.~\cite{Cea98}), the deconvolved FWHM of these sources are shown as dashed
ellipses, which may be partially outside the plotted area.}
\end{figure*}

\begin{figure*}
\centerline{\psfig{figure=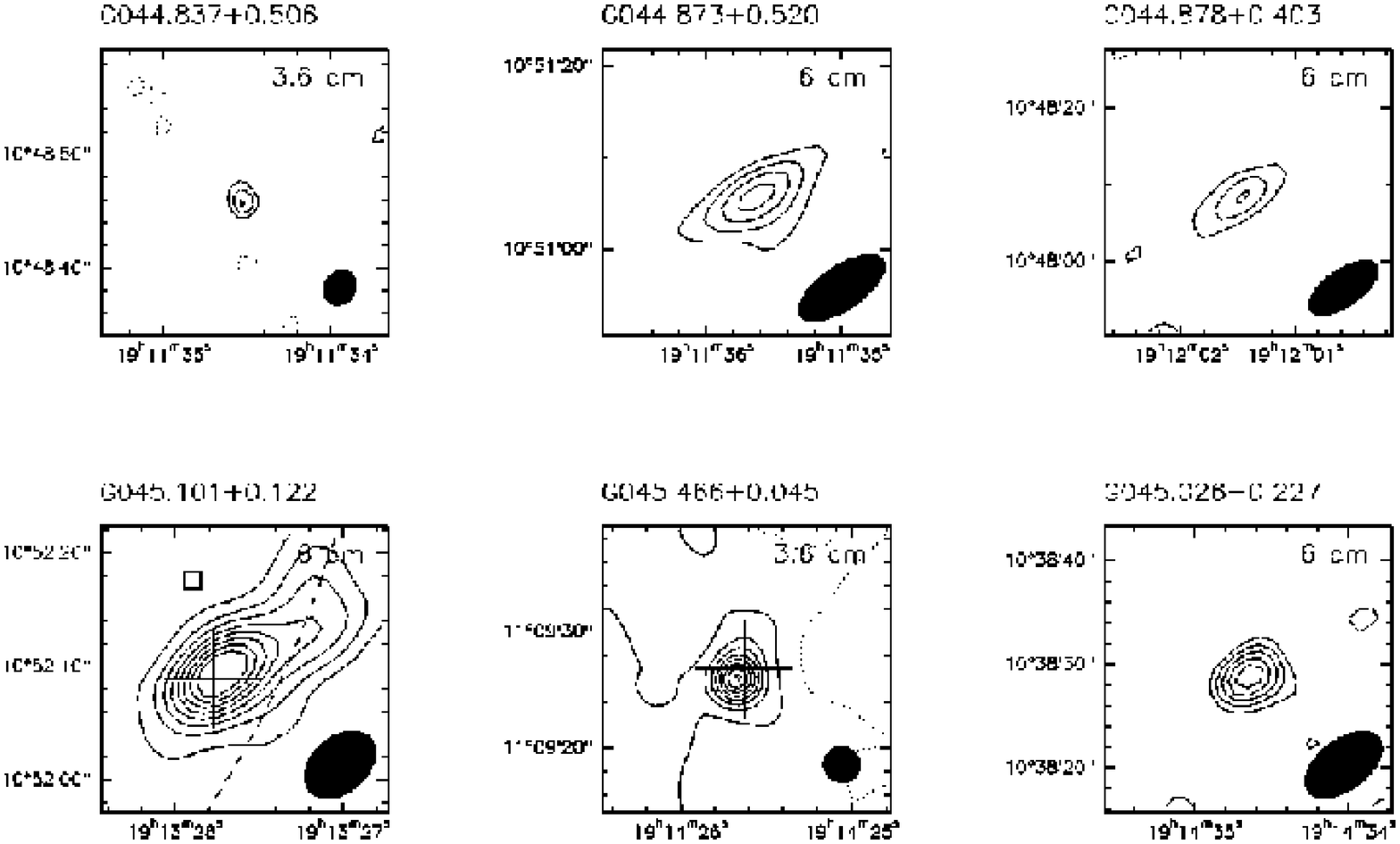,width=14.2cm,angle=-90}}
\vskip 0.9cm
\centerline{\psfig{figure=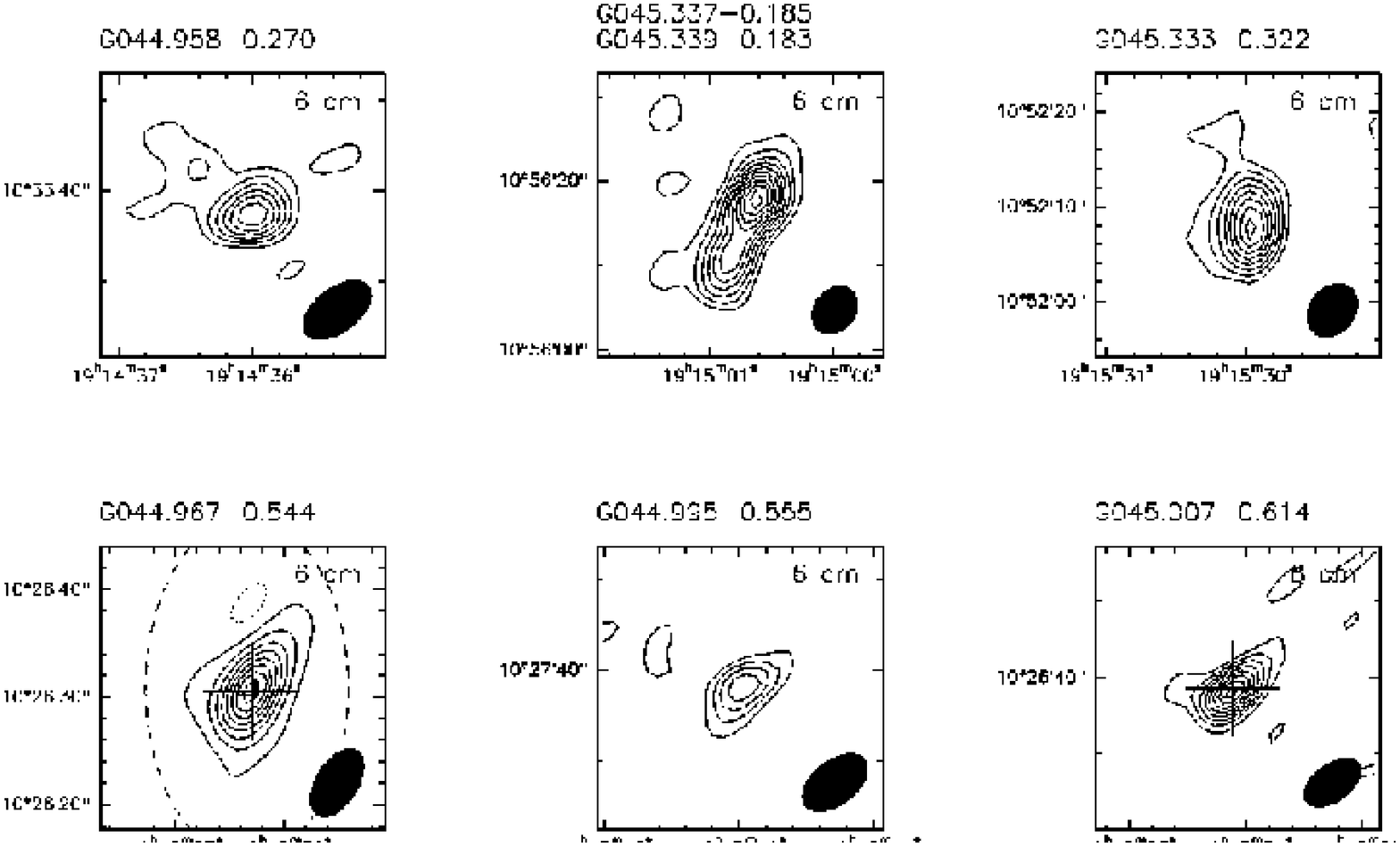,width=14.2cm,angle=-90}}
\vskip 0.4cm
\caption[]{\label{pcfig2}As Fig.~\ref{pcfig1} but for sources
detected only at one frequency.}

\end{figure*}

In this appendix we present the contour plots of all the sources
detected in our survey.
IRAS-PSC2 sources are shown as ellipses corresponding to the 
90\% confidence on the peak position. The positions of the peaks of
the 20~cm sources from Zoonematkermani et al.~(\cite{Zea90})
are shown as crosses. Dashed ellipses show the deconvolved sizes
of the 20~cm NVSS survey sources (Condon et al.~\cite{Cea98}).
In Fig.~\ref{pcfig1} and~\ref{pcfig2} the NVSS ellipses may fall
partially (or completely) outside the plotted area, thus we marked
the NVSS peak position with an empty square.

%
%
%

\begin{figure*}
\centerline{\psfig{figure=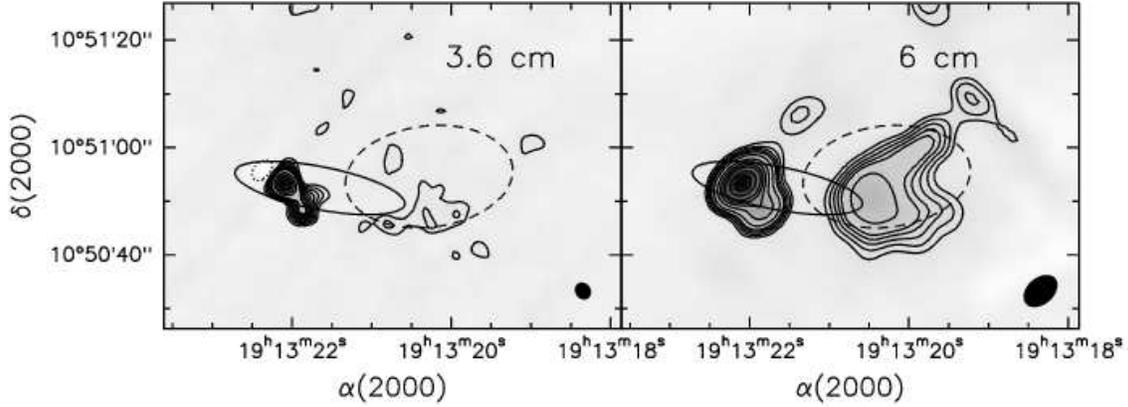,width=15cm,angle=-90}}
\caption[]{\label{exfig1}3.6 and 6~cm maps of the extended source
G045.066$+$0.138 (30), and the compact sources G045.070$+$0.132 (11) and
G045.072$+$0.132 (12); contour levels
are -8, from 8 to 50 by 5 and from 50 to 300 by 50~mJy/beam at 3.6~cm, and
-10, from 10 to 22 by 3, from 30 to 90 by 15, 120 and 150~mJy/beam at 6~cm.
The filled ellipse represent the IRAS-PSC error box.}
\end{figure*}

\begin{figure*}
\centerline{\psfig{figure=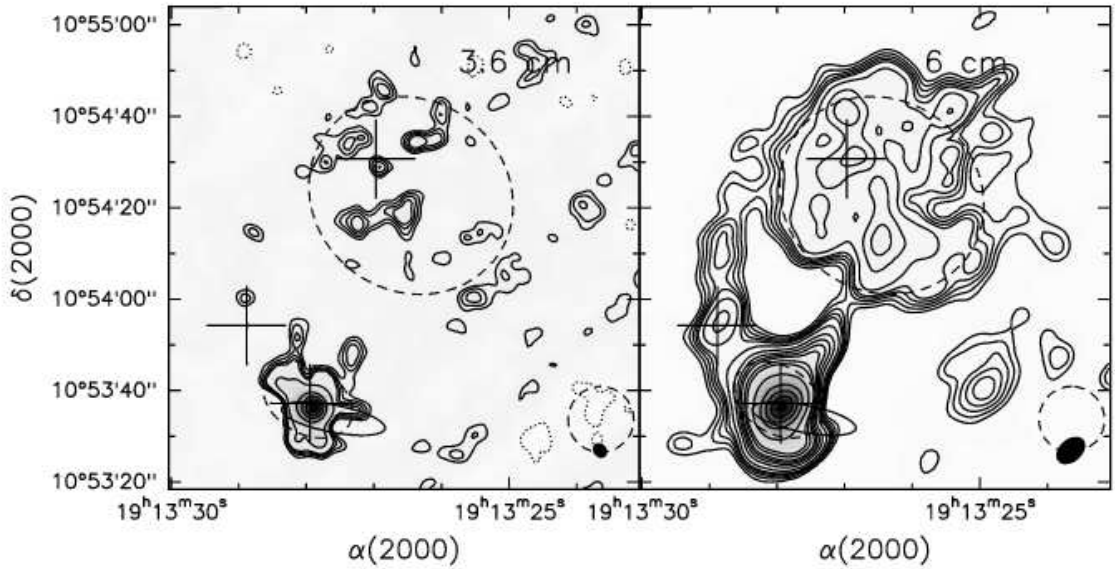,width=15cm,angle=-90}}
\caption[]{\label{exfig2}3.6 and 6~cm maps of the extended source
G045.134$+$0.145 (31), and the compact sources G045.118$+$0.143 (13),
G045.123$+$0.132 (15), G045.133$+$0.133 (16) and G045.130$+$0.131 (17);
contour levels
are $-$30, from 30 to 100 by 10 and from 200 to 1400 by 300~mJy/beam at 3.6~cm,
and
$-$10, from 10 to 22 by 3, from 30 to 90 by 15, 120, 150 and from
300 to 1500 by 300~mJy/beam at 6~cm.
The filled ellipse represent the IRAS-PSC error box.}
\end{figure*}

\begin{figure*}
\centerline{\psfig{figure=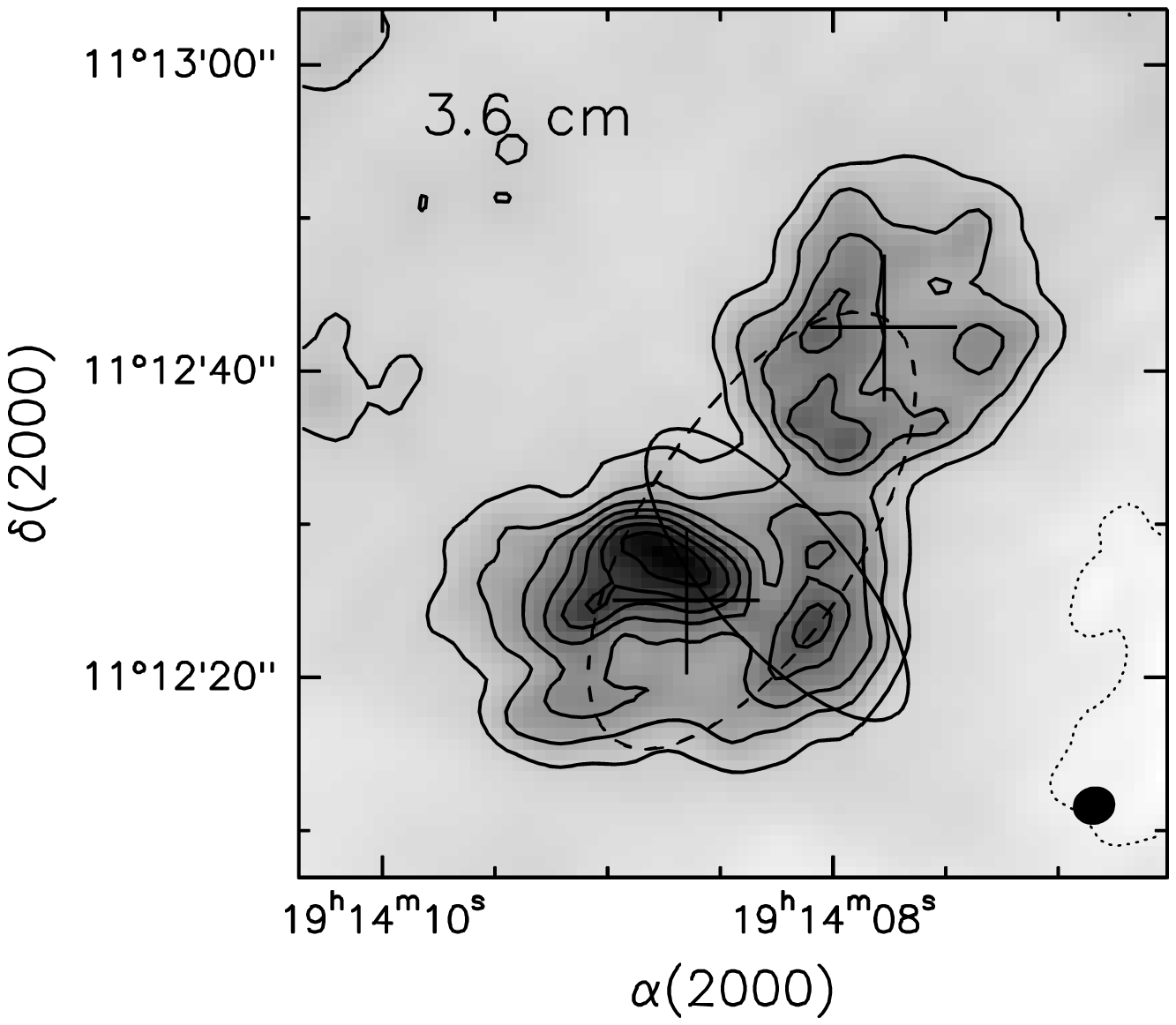,width=7.8cm,angle=-90}
            \hskip 0.5cm
            \psfig{figure=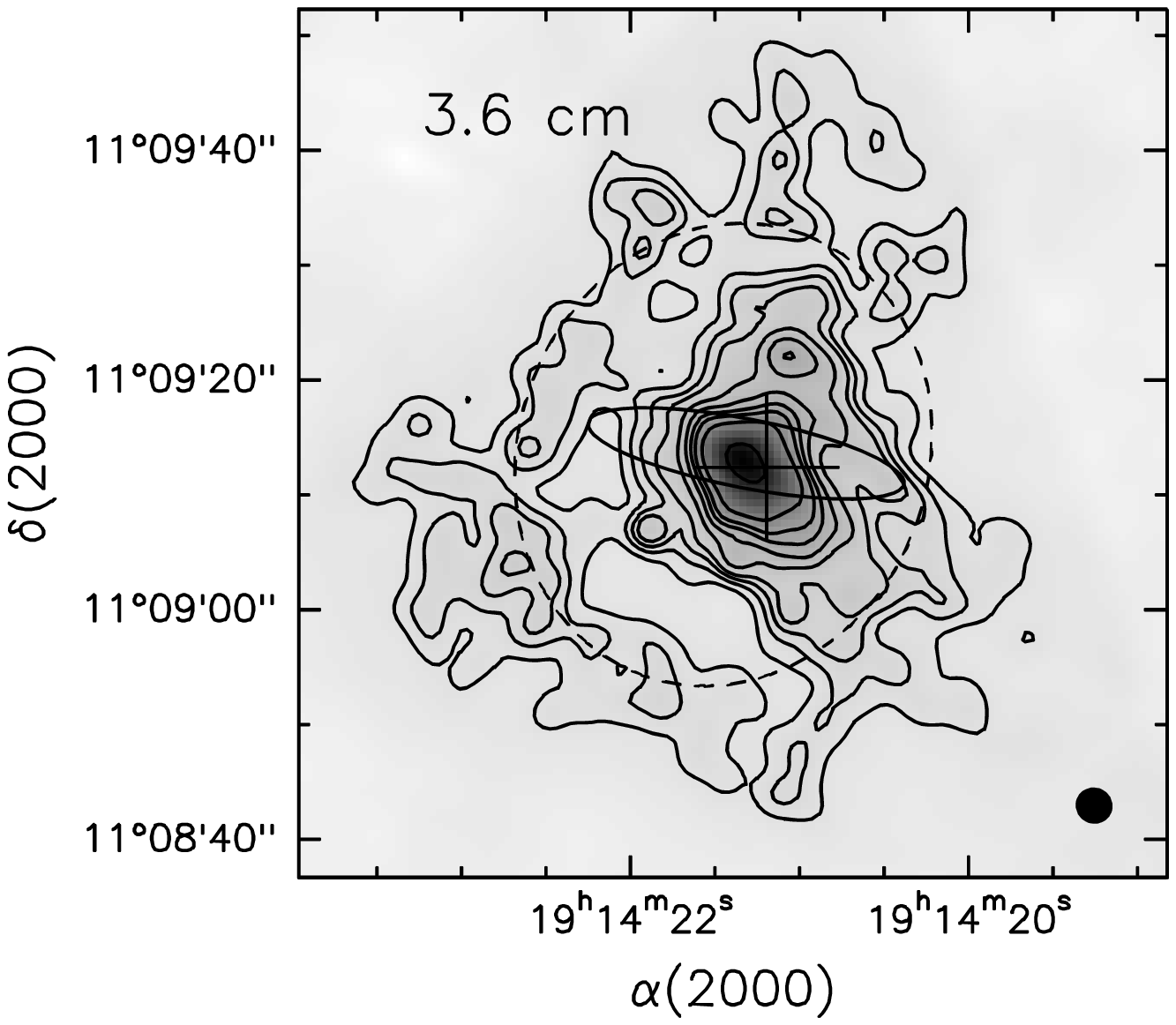,width=7.8cm,angle=-90}}
\caption[]{\label{exfig3}3.6~cm maps of the extended sources
G045.479$+$0.130 (32) on the left panel and 045.455$+$0.059 (33)
and compact source G045.455$+$0.060 (18) on the right panel;
contour levels are:
$-3$, from 3 to 37 by 5~mJy/beam, and 
$-$9, from 9 to 27 by 6, from 39 to 75 by 12, 100, 150 and 200~mJy/beam.
The filled ellipse represent the IRAS-PSC error box.}
\end{figure*}

\begin{figure*}
\centerline{\psfig{figure=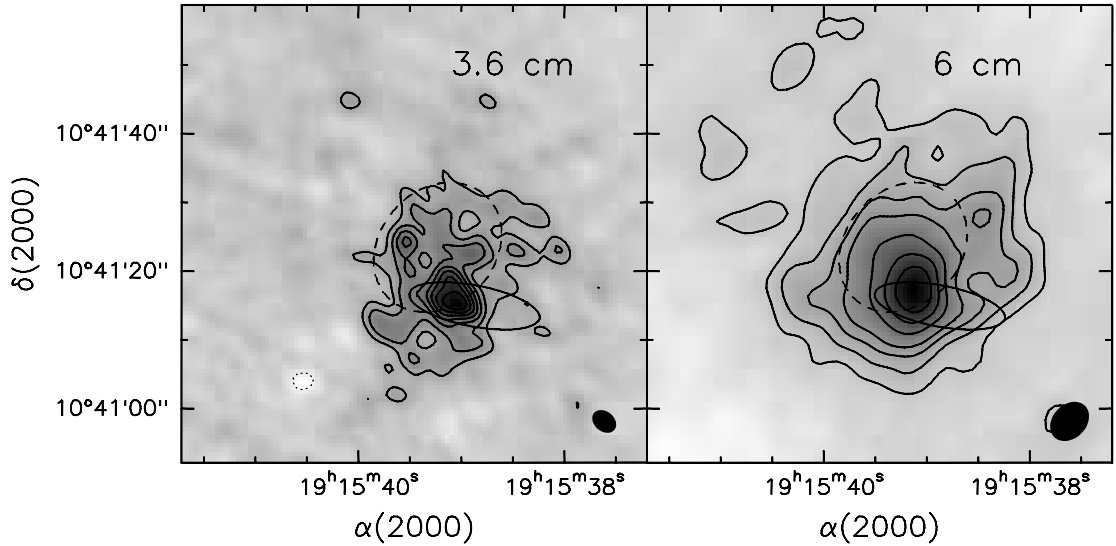,width=15cm,angle=-90}}
\caption[]{\label{exfig4}3.6 and 6~cm maps of the extended source
G045.190$-$0.439 (34); contour levels
are -0.6, from 0.6 to 3.3 by 0.4~mJy/beam at 3.6~cm, and
-0.9, from 0.9 to 2.7 by 0.6, from 3.9 to 7.5 by 1.2~mJy/beam at 6~cm.
The filled ellipse represent the IRAS-PSC error box.}
\end{figure*}

\end{document}